\documentclass[10pt,aps,prd,reprint,nofootinbib,floats,floatfix,amsfonts,amssymb,amsmath,preprintnumbers,notitlepage,superscriptaddress,twocolumn,longbibliography]{revtex4-2}

\newcommand{\AxisRotator}[1][rotate=112]{%
    \tikz [x=0.15cm,y=0.40cm,line width=.2ex,-stealth,#1] \draw (0,0) arc (-150:150:1 and 1);%
}

\usepackage{url}
\usepackage{graphicx,color} 
\usepackage{hyphenat} 
\usepackage{amsmath, amssymb, amsfonts,amsbsy,amsthm} 
\usepackage[inkscapeformat=png]{svg}
\usepackage{float}
\usepackage{subcaption}
\usepackage{soul,xcolor} 
\usepackage{tikz}
\usepackage{pgfpages}
\usepackage{tikz-3dplot}
\usepackage{hyperref}
\usepackage{tabularx}
\usepackage{multirow}
\usepackage{booktabs}
\usepackage{siunitx}
\usepackage{amsfonts}
\usepackage{bm}
\usetikzlibrary{shapes.geometric, arrows}
\usepackage{pgfplots,tikz-3dplot}
\tikzstyle{startstop} = [rectangle, rounded corners, 
minimum width=3cm, 
minimum height=1cm,
text centered, 
draw=black, 
fill=red!30]

\tikzstyle{io} = [trapezium, 
trapezium stretches=true, 
trapezium left angle=70, 
trapezium right angle=110, 
minimum width=3cm, 
minimum height=1cm, text centered, 
draw=black, fill=blue!30]

\tikzstyle{process} = [rectangle, 
minimum width=3cm, 
minimum height=1cm, 
text centered, 
text width=3cm, 
draw=black, 
fill=orange!30]

\tikzstyle{vetos} = [rectangle, 
minimum width=3cm, 
minimum height=1cm, 
text centered, 
text width=3cm, 
draw=black, 
fill=red!30]

\tikzstyle{detectors} = [rectangle, 
minimum width=3cm, 
minimum height=1cm, 
text centered, 
text width=3cm, 
draw=black, 
fill=green!30]

\tikzstyle{arrow} = [thick,->,>=stealth]

\usetikzlibrary{calc}
\setstcolor{red}

\newcommand{\orcid}[1]{\href{https://orcid.org/#1}{\includegraphics[scale=0.15]{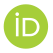}}}

\newcommand{\MDGold}[1]{{#1}}
\newcommand{\MDG}[1]{{#1}}
\usepackage[normalem]{ulem}

\begin{document}

\title{Doppler correlation-driven vetoes for the Frequency Hough analysis in continuous gravitational-wave searches}
\author{Matteo Di Giovanni\, \orcid{0000-0003-4049-8336}}\email[]{matteo.digiovanni@uniroma1.it}
\affiliation{La Sapienza Università di Roma, I-00185 Roma, Italy}\affiliation{INFN, Sezione di Roma, I-00185 Roma, Italy}
\author{Paola Leaci\, \orcid{0000-0002-3997-5046}}\email[]{paola.leaci@uniroma1.it} 
\affiliation{La Sapienza Università di Roma, I-00185 Roma, Italy}\affiliation{INFN, Sezione di Roma, I-00185 Roma, Italy}
\author{Pia Astone \orcid{0000-0003-4981-4120}}
\affiliation{INFN, Sezione di Roma, I-00185 Roma, Italy}

\author{Stefano Dal Pra \orcid{0000-0002-1057-2307}}
\affiliation{INFN - CNAF, I-40129 Bologna, Italy}
\author{Sabrina D'Antonio \orcid{0000-0003-0898-6030}}
\affiliation{INFN, Sezione di Roma, I-00185 Roma, Italy}
\author{Luca D'Onofrio \orcid{0000-0001-9546-5959}}
\affiliation{INFN, Sezione di Roma, I-00185 Roma, Italy}
\author{Sergio Frasca}
\affiliation{INFN, Sezione di Roma, I-00185 Roma, Italy}
\author{Federico Muciaccia \orcid{0000-0003-0850-2649}}
\affiliation{La Sapienza Università di Roma, I-00185 Roma, Italy}
\affiliation{INFN, Sezione di Roma, I-00185 Roma, Italy}
\author{Cristiano Palomba \orcid{0000-0002-4450-9883}}
\affiliation{INFN, Sezione di Roma, I-00185 Roma, Italy}
\author{Lorenzo Pierini \orcid{0000-0003-0945-2196}}
\affiliation{INFN, Sezione di Roma, I-00185 Roma, Italy}
\author{Francesco Safai Tehrani \orcid{0000-0001-7796-0120}}
\affiliation{INFN, Sezione di Roma, I-00185 Roma, Italy}

\date{\today}

\begin{abstract}
We present an improved method for vetoing candidates of continuous gravitational-wave sources during all-sky searches utilizing the Frequency Hough pipeline. This approach leverages linear correlations between source parameters induced by the Earth Doppler effect, which can be effectively identified through the Hough Transform. Candidates that do not align with these patterns are considered spurious and can thus be vetoed, enhancing the depth and statistical significance of follow-up analyses. 
Additionally, we provide a comprehensive explanation of the method calibration, which intrinsically linked to the total duration of the observing run. On average, the procedure successfully vetoes $56\%$ of candidates. 
To assess the method performance, we conducted a MC simulation injecting fake continuous-wave signals into data from the third observing run of the LIGO detectors. This analysis allowed us to infer strain amplitude upper limits at a $90\%$ confidence level. We found that the optimal sensitivity is $h_0^{90\%} = 3.62^{+0.23}_{-0.22}\times 10^{-26}$ in the [128, 200] Hz band, which is within the most sensible frequency band of the LIGO detectors.

\end{abstract}

\maketitle


\section{Introduction}\label{sec:intro}

Over the last decade, the Advanced LIGO (aLIGO) \cite{aLIGO} and Advanced Virgo (aVIRGO) \cite{aVirgo} detectors successfully carried out three observing runs (O1, O2 and O3)\cite{GWOSC1,GWOSC}, with a fourth run (O4) still ongoing and expected to collect a total of more than two years of data taking. During the first three observing runs, aLIGO and aVirgo made numerous transient gravitational-wave (GW) detections, coming from the inspirals and mergers of compact binary systems of black holes (BH) \cite{gwtc1, gwtc2, gwtc3}, neutron stars (NS) \cite{gwtc1, gwtc2, gwtc3}, as well as NS-BH binaries \cite{gwtc3}. However, several other forms of GW radiation have yet to be detected. Among these are long-lasting, nearly monochromatic continuous waves (CW) emitted by rotating, non-axisymmetric NSs, with GW frequencies close to or related to their spin frequency \cite{Lasky_2015, WETTE2023102880}. Additionally, the frequency changes much more slowly compared to transient sources, occurring over timescales of years rather than seconds.

Since the CW emission mechanism is closely tied to the internal properties and evolution of NSs, detecting CWs from these stars would provide valuable insights into their enigmatic interiors \cite{WETTE2023102880}. It would also offer a chance to study dense matter under conditions different from those present in binary NS inspirals and mergers, and allow for additional tests of gravitational theory \cite{isi}. Due to the inherently weaker GW amplitude of CWs compared to the already-detected transient sources, searches for CWs from rotating non-axisymmetric NSs are primarily confined to the Milky Way and require extended observation periods to improve the signal-to-noise ratio.

\MDGold{Within the LIGO-Virgo-KAGRA collaboration (LVK)}, in the context of CW investigations, there are several pipelines dedicated to all-sky searches \cite{CW1,CW2,CW5} for isolated NSs \cite{Astone2014, skyhough, fstat1, fstat2, soap}, with these being among the most computationally intensive GW searches. A \MDGold{comprehensive} review of pipelines for wide parameter-space searches\MDGold{, including those used outside the LVK collaboration,} can be found in \cite{universe7120474}. Among these pipelines is the Frequency Hough (FH) method \cite{Astone2014}, which is the focus of this work. This method employs the Hough transform (HT) \cite{hough59, hough} to reconstruct the source parameters.

All sky searches, which explore a wide range of the parameter space, usually return a huge number of source candidates that are passed to follow up and post-processing procedures to increase the detection statistics. Since a substantial number of candidates is often due to noise, various veto procedures are continually developed and tested to discard false positives and increase the robustness and statistical significance of follow-up analyses. \MDGold{One way to distinguish candidates genuinely associated with a source from noise is by considering the Doppler modulation of the GW signal, which arises from the relative motion of the detector with respect to the source. It is highly unlikely that a noise artifact would exhibit this modulation, which is caused by the Earth orbital motion \cite{Isiveto}. Building on this idea, several veto strategies \cite{zhu,Isiveto,Intini,gionz}, all complementary to each other and applicable to different search methods, have been developed. Some approaches \cite{zhu,Isiveto,gionz} investigate the impact of Doppler modulation on the signal morphology by turning it on and off during the signal reconstruction process to assess the statistical significance of candidates. Other methods \cite{Intini} exploit correlations among candidate parameters \cite{Prix_2005, pletsch}, which arise from the Doppler effect induced by Earth's motion during the observation period.}

In this work, we build upon the set of vetoes introduced by \cite{Intini} and specifically designed to be integrated in FH pipeline procedures. 
Since the Doppler induced correlations manifest as distinct patterns in the parameter space, we apply the HT to the FH candidates to identify them. Originally, the HT was developed for detecting patterns in particle trajectories in bubble chamber analysis \cite{hough59} and later adapted for feature extraction in digital image analysis and processing.

Based on \cite{Intini}\MDGold{, which has already demonstrated the impact of this method on the performance of the FH pipeline,} we refine the methodology by providing a more detailed description and a comprehensive explanation of the calibration of the veto procedure. Additionally, we conduct a Monte Carlo (MC) simulation to determine the upper limits on the amplitude at a 90~$\%$ confidence level (CL) for the procedure.

The paper is organized in the following way: in Section \ref{sec:signal_model} we briefly summarize the properties of CW signals from isolated NS; Section \ref{sec:search_methods} is devoted to an overview of the FH pipeline, whereas the veto procedure and its practical implementation is described in Sections \ref{sec:vetoes} and \ref{sec:procedure}, respectively. The efficiency of the procedure is shown in Section \ref{sec:efficiency} and discussed in Section \ref{sec:conc}.

\section{Signal Model}
\label{sec:signal_model}

 In the detector frame, the CW signal from an isolated, \MDGold{non-axisymmetric} NS spinning around one of its principal axis of inertia is given by \cite{jaranowski2, Lasky_2015, WETTE2023102880}
 
 \begin{equation*}
     h(t)=h_0 [F_+(t,\lambda, \beta, \psi)\frac{1+\cos^2\iota}{2}\cos\phi(t)+
 \end{equation*}
 \begin{equation}\label{eq:sig_model}
     +F_{\times}(t,\lambda, \beta, \psi)\cos\iota\sin\phi(t)],
 \end{equation}
where $F_+$ and $F_{\times}$ are the antenna patterns of the detectors dependent on time $t$, \MDGold{ecliptic} longitude $\lambda$ and \MDGold{ecliptic} latitude $\beta$ of the source \MDGold{(referred to simply as longitude and latitude henceforth)} and polarization angle $\psi$; $h_0$ is the signal amplitude; $\iota$ is the angle between the star total angular momentum vector and the direction from the star to the Earth; $\phi(t)$ is the signal phase. Assuming a source at distance $d$ emitting CWs at twice the rotational frequency, the signal amplitude can be written as
\begin{equation*}
     h_0 = \frac{4\pi^2G}{c^4}\frac{\epsilon I_{zz}f^2}{d}\simeq1.06\times 10^{-26}\left ( \frac{\epsilon}{10^{-6}}\right )\times 
\end{equation*}
\begin{equation}\label{eq:sig_amp}
    \times \left ( \frac{I_{zz}}{10^{38}\rm\,{kg/m^2}}\right )\left ( \frac{f}{100 \rm\,{Hz}}\right )^2\left ( \frac{1\rm\,kpc}{d}\right ),
\end{equation}
with $\epsilon$ being the NS ellipticity and $I_{zz}$ the moment of inertia of the star with respect to the principal axis (aligned with the rotation axis).

In the detector frame, the received frequency $f(t)$ does not correspond to a simple monochromatic signal. 
Initially, $f(t)$ is Doppler-shifted due to Earth  motions and is related to the emitted frequency $f_0(t)$ by \cite{Astone2014, WETTE2023102880}
\begin{equation}\label{eq:standard_doppler0}
    f(t)=f_0(t) \left (1 + \frac{\vec{v}\cdot \hat{n}}{c} \right ),
\end{equation}
where the term $\frac{\vec{v}\cdot \hat{n}}{c}$ identifies the Doppler shift experienced by a CW due to Earth motions. Here, $\vec v=\vec{v}_{\textrm{rev}} + \vec{v}_{\textrm{rot}}$, denotes the total velocity of the motion of the Earth, \MDGold{which is given by the combination of the revolution around the Sun (denoted by the subscript "rev") and the rotation around its axis (denoted by the subscript "rot")}. 
$\hat n$ identifies the direction of the source with respect to the detector. The maximum Doppler shift for a given source $f_0$ is $f_0|\vec v|/c\simeq 10^{-4}f_0$.
Additionally, the intrinsic signal frequency $f_0(t)$ varies over time as the NS gradually slows down due to the loss of rotational energy through the emission of electromagnetic and gravitational radiation. Hence, $f_0(t)$ can be expressed as \cite{Astone2014, WETTE2023102880}
\begin{equation}\label{eq:sd_def}
    f_0(t)=f_0 + \dot f_0(t-t_0),
\end{equation}
where $\dot f_0$ is the first time derivative of the frequency and represents the spin down parameter.
In general, CW from isolated NSs can be described by $8$ parameters: frequency $f_0$; spin down $\dot f_0$; source sky location (longitude $\lambda$ and latitude $\beta$); inclination $\iota$; polarization angle $\psi$;  phase $\phi$; signal amplitude $h_0$. Please note that the frequency evolution $f_0(t)$ of the source is distinct from the frequency parameter $f_0$ which is $f_0=f_0(t=t_0)$ \cite{WETTE2023102880}.

\section{The Frequency Hough pipeline}
\label{sec:search_methods}
The FH pipeline \cite{Astone2014} is a semi-coherent method for reconstructing CW source parameters using the HT \cite{hough, hough59}.
In this approach, the data from the entire observation period of the observing run ($T_{\textrm{obs}}^{\textrm{tot}}$) is divided into shorter segments, leading to a loss of phase and polarization information. Consequently, the number of source parameters is reduced from eight to five $(h_0,f,\dot f, \lambda, \beta)$ \cite{Astone2014}. Then, each segment is analyzed coherently through Fast Fourier Transform (FFT), reflecting the semi-coherent nature of the pipeline. Despite the sensitivity loss compared to a fully coherent search, which is computationally more demanding, this loss has been found to be minimal~\cite{Astone2014}.
The FH pipeline has been employed in several all-sky searches using Virgo and LIGO data \cite{CW1, CW2, CW3, CW4, CW5}. While a comprehensive description of the method is outside the scope of this work, it can be found in \cite{Astone2014}. Here, we provide a brief overview of the main analysis steps and the parameters used.

The duration $\rm T_{FFT}$ of the FFT depends on the frequency band being considered (see Table \ref{tab:FH}) and is such that, if a signal is present, the frequency spread caused by the Doppler effect remains smaller than the frequency bin width $\delta f = 1/\rm T_{FFT}$. Additionally, $\rm T_{FFT}$ defines the spin-down bin size, which is $\delta\dot f=\delta f/\rm T_{\textrm{obs}}^{\textrm{tot}}$, where $T_{\textrm{obs}}^{\textrm{tot}}$ is the total observing run duration.

\begin{table}[h!]

\begin{tabular}{cccc}
Band [Hz] & $\rm\, T_{FFT}$ [s] & $\delta f$ [Hz] & $\delta\dot f$ [Hz/s] \\ 
\hline
$10-128$ & $8192$ & $1.22\times10^{-4}$ & $3.92\times10^{-12}$\\ 
$128-512$ & $4096$ & $2.44\times10^{-4}$ & $7.83\times10^{-12}$\\ 
$512-1024$ & $2048$ & $4.88\times10^{-4}$ & $1.57\times10^{-11}$\\ 
$1024-2048$ & $1024$ & $9.76\times10^{-4}$ & $3.13\times10^{-11}$\\ 
\hline 
\end{tabular}

\caption{Properties of the FFTs used for the FH pipeline. The time duration $\rm T_{FFT}$ refers to the length in seconds of the data chunks on which the FFT is computed. The frequency bin width is $1/\rm T_{FFT}$, whereas $\delta\dot f=\delta f/\rm T_{\textrm{obs}}^{\textrm{tot}}$, where $\rm T_{obs}$ is the total run duration.}
\label{tab:FH}
\end{table}
 

For each FFT, the procedure calculates the ratio between its squared modulus and an auto-regressive estimate of the average power spectrum \cite{Astone_2005}. The most significant peaks in the time-frequency domain are selected setting a threshold $\Theta=1.58$ on this ratio, which corresponds to a false alarm probability of $P_{FA} = 0.01$ \cite{Astone2014, CW5}. This collection of peaks forms the so-called peakmap \cite{Astone2014}, which is then cleaned of the strongest disturbances using a line persistency veto. Subsequently, for each point on the sky grid, the time-frequency peaks in the peakmap are adjusted to account for the Doppler effect due to the motion of the detector.

The shifted peaks are then fed to the FH algorithm, which transforms each peak to the frequency-spin down plane of the source. The output is a 2-D histogram in $(f,\dot f)$ \cite{Astone2014}. Significant points in this plane are identified by dividing each 1 Hz band of the corresponding histogram into 20 intervals. For each interval and each sky location, the candidates with the highest statistical significance are selected. Typically, this involves choosing one or, in most cases, two candidates with the highest critical ratio (CR), which measures their statistical significance \cite{CW5}
\begin{equation}
    CR = \frac{x-\mu}{\sigma},
\end{equation}
where $x$ is the value of the peakmap projection in a given frequency bin; $\mu$ is the average value; $\sigma$ is the standard deviation of the peakmap projection.
All the steps described so far are applied separately to each detector involved in the analysis.

Coincident candidates among the pair of detectors are found using a distance metric built in the four-dimensional parameter space of sky position $(\lambda, \beta)$ (longitude and latitude), frequency $f$ and spin down $\dot f$, i.e.:
\begin{equation}\label{eq:distance}\scriptstyle
    d=\sqrt{\left ( \frac{f_1-f_2}{\delta f} \right )^2 + \left ( \frac{\dot f_1-\dot f_2}{\delta \dot f} \right )^2 + \left ( \frac{\lambda_1-\lambda_2}{\delta \lambda} \right )^2 + \left ( \frac{\beta_1-\beta_2}{\delta \beta} \right )^2},
\end{equation}
where $\delta f$,  $\delta \dot f$,  $\delta \lambda$,  $\delta \beta$ are the parameter bin widths, while the labels 1 and 2 refer to the pair of candidates being considered for the coincidence step. Pairs of candidates with distance $d \le 3$ are considered coincident. 

At this point, the most significant coincident candidates are selected and subject to five follow up steps \cite{CW5} to discard false candidates and increase the detection confidence. Since the veto procedure described in the current work is designed to be applied after the coincidence between detectors, but before any post processing, the follow up will not be discussed here.

\subsection{{Data set used}}

The data set used in this analysis is the publicly available O3 data set of the aLIGO GW detectors \cite{GWOSC}. We decide to not consider the Virgo detector, since its sensitivity is significantly worse and would negatively affect our results. LIGO is made up of two laser interferometers, both with 4 km long arms. One is at the LIGO Livingston Observatory (L1) in Louisiana, USA and the other is at the LIGO Hanford Observatory (H1) in Washington, USA. The O3 run lasted from 2019 April 1 and 2020 March 27. The run was divided into two parts, O3a and O3b, separated by one month commissioning break that took place in October 2019. The duty factors for this run were $76\%$, $71\%$, for L1, H1 respectively. The maximum uncertainties ($68\%$CL) on the calibration of the LIGO data were of $ 7 \% / 11 \%$ in amplitude and 4 deg/9 deg in phase for O3a/O3b data \cite{o3char}.

\section{Doppler-based correlations}
\label{sec:vetoes}

Combining Equations \ref{eq:standard_doppler0} and \ref{eq:sd_def}, the frequency evolution in the detector frame of the CW signal emitted by an isolated NS is 
\begin{equation}\label{eq:standard_doppler}
    f(t)=(f_0+\dot f_0t) \left (1 + \frac{\vec{v}\cdot \hat{n}}{c} \right ),
\end{equation}
where the parameters are the same as those defined in Section \ref{sec:signal_model}.
\tdplotsetmaincoords{70}{110}
\begin{figure}[h]
\begin{tikzpicture}[scale=0.65, z={(0,1cm)},
x={(-3.85mm,-3.85mm)},
y={(1cm,0)},
thick,scale=2]
\begin{scope}[->,ultra thick]
\draw(0,0,0) -- (3,0,0);
\draw (0,0,0) -- (0,3,0);
\draw (0,0,0) -- (0,0,2);
\end{scope}
\draw (0,-1,3) node[left] {\textbf{SSB}};
\shade[ball color=orange, opacity=1] (0,0) circle (0.3cm);
\draw(0,0,0.5) node[left = 3pt]{\textbf{Sun}};
\draw [->] (0,0,0) -- (1.2,1.3,4);
\draw (2,1.6,3) node[]{$\hat n_{\textrm{SSB}}$};
\draw[dashed] (0,0) ellipse (3cm and 1cm);
\begin{scope}[rotate=20, xscale=0.2, yscale=0.2, shift={(-60,1)}]
      \coordinate (O) at (0,0,0);
      \shade[ball color=cyan, opacity=1] (O) ellipse (1cm and 0.66cm);
      \coordinate(Om) at (0,0,2.5);
      \node[above] at (0,0,2.7) {$f_0, \dot f_0$};
      \node[above] at (0,0.1,4.5) {NS at $(\lambda_0, \beta_0)$};
      \draw[->, line width=0.5mm, fill = black] (O) -- (Om) node at (0,0,1.2) {\AxisRotator};
    \end{scope}
\begin{scope}[shift={(2.3,2.3)},
]
\draw[->] (0,0,0) -- (0,-0.2,0.5) node [very near end] {\AxisRotator} node[above = 3pt]{$\vec v_{\textrm{rot}}$};
\shade[ball color=blue, opacity=1] (0,0) circle (0.2cm);
\draw (0,0,-0.35) node[below]{\textbf{Earth}};
\draw[->] (0,0,0) -- (-0.4,0.5,0) node[below=6pt]{$\vec v_{\textrm{rev}}$};
\draw[-] (0,0,0) -- (-2.3,-2.3);
\draw (-1,-1) node[left=3pt]{$R$};
\draw [->] (0,0,0) -- (1.2,1.4,4);
\draw (2,1.6,3) node[above]{$\hat n$};
\end{scope}
\end{tikzpicture}
\caption{Representation of a source rotating at frequency $f_0$ with spindown $\dot f$ in the SSB. For sources far away from the Earth, $\hat n$ and $\hat n_{\textrm{SSB}}$ can be assumed as being parallel.}
\label{fig:SSB}
\end{figure}
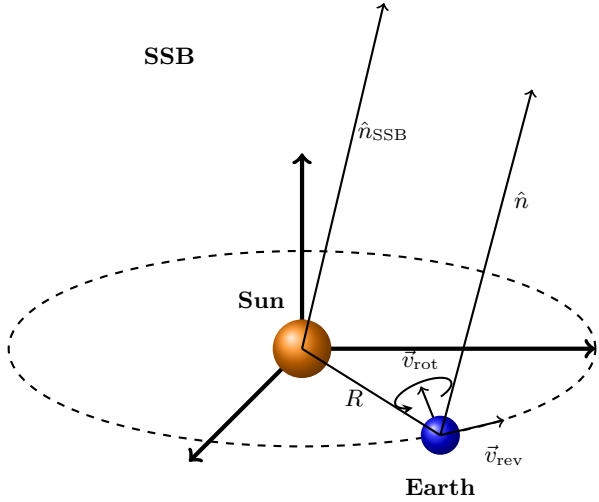
In particular, in the solar system barycentre  reference frame (SSB, Figure \ref{fig:SSB}), $\vec{v}=\vec{v}_{\textrm{rev}} + \vec{v}_{\textrm{rot}}$ can be written as \cite{jotania}

\begin{equation}\label{eq:velocities}
    \vec{v} = \begin{pmatrix}-R\omega_{\textrm{rev}}\sin(\omega_{\textrm{rev}} t) - r\omega_{\textrm{rot}}\sin\alpha\sin(\omega_{\textrm{rot}} t) \\ 
    R\omega_{\textrm{rev}}\cos(\omega_{\textrm{rev}} t) + r\omega_{\textrm{rot}}\sin\alpha\cos(\omega_{\textrm{rot}} t)\cos\gamma \\ 
    r\omega_{\textrm{rot}}\sin\alpha\sin(\omega_{\textrm{rot}} t)\sin\gamma\end{pmatrix}
\end{equation}

with $\alpha, \gamma$ \MDGold{being the equatorial longitude and latitude that define} the position of the detector on the Earth's surface and $r$ is the Earth radius to keep into account the finite size of the Earth; $R$ is the Sun-Earth distance; $\omega_{\textrm{rot}}$ and $\omega_{\textrm{rev}}$ are the Earth rotation and orbital angular velocities, respectively. From equation \ref{eq:velocities}, we note that the factor $r\omega_{\textrm{rot}} = 0.46 \rm\, km/s$ is $10^{-2}$ times $R\omega_{\textrm{rev}} = 29.78 \rm\, km/s$ and we neglect it for simplicity \cite{Intini, gionz};
this is equivalent of retaining only the velocity component of the orbital motion of the Earth:
\begin{equation}\label{eq:standard_doppler_2}
    f=(f_0+\dot f_0t) \left (1 + \frac{\vec{v}_{\textrm{rev}}\cdot \hat{n}}{c} \right ) + \mathcal{O}(\omega_{\textrm{rot}}),
\end{equation}
with
\begin{equation}
    \vec v_{\textrm{rev}} = \left [ -R\omega_{\textrm{rev}}\sin(\omega_{\textrm{rev}} t); R\omega_{\textrm{rev}}\cos(\omega_{\textrm{rev}} t); 0 \right ].
\end{equation}
To simplify the notation, from now on we will refer to $\vec v_{\textrm{rev}}$ as $\vec v$, to $|\vec v|$ as $v$ and to $\omega_{\textrm{rev}}$ as $\omega$. Moreover, since the source is assumed to be far from the detector (Figure \ref{fig:SSB}), its distance being much larger than the average distance between the detector and the barycentre of the Solar System, $\hat n$ can be assumed parallel to the unit vector drawn from the centre of the SSB to the source, which is  
\cite{WETTE2023102880,Srivastava}
\begin{equation}
    \hat{n}_{\textrm{SSB}}\equiv \left [ \sin(\beta'_0)\cos(\lambda_0);\sin(\beta'_0)\sin(\lambda_0);\cos(\beta'_0)\right ],
\end{equation}
where $(\beta'_0,\lambda_0)$ are the colatitude $\beta_0'\equiv 90-\beta_0$ and the longitude of the source in heliocentric ecliptic coordinates. Therefore, Eq. \ref{eq:standard_doppler_2} becomes
\begin{equation}\label{eq:doppler_line}
    f=(f_0+\dot f_0t)\left[ 1 + \frac{v}{c}\sin\beta'_0\sin(\lambda_0 - \omega t) \right ].
\end{equation}
Equation \ref{eq:doppler_line} can be further simplified. In fact, we observe that $\frac{v}{c}\simeq 10^{-4}$ and, for spinning isolated NS, $\dot f_0 < 10^{-9} {\rm \frac{Hz}{s}}$. Moreover, the product of the trigonometric function oscillates between $-1$ and $1$. Therefore, considering, e.g., $T_{\textrm{obs}}=\mathcal{O}( 1 \rm\,{yr}) \simeq 3\times 10^{7} \rm\, s$, the term $\dot f_0t\frac{v}{c}\sin\beta'_0\sin(\lambda_0 - \omega t)\simeq \mathcal{O}( 10^{-6})$ can be neglected with respect to the term in $f_0$ and Equation \ref{eq:doppler_line} can be rewritten as

\begin{equation}\label{eq:doppler_line_2}
    f=f_0\left[ 1 + \frac{v}{c}\sin\beta'_0\sin(\lambda_0 - \omega t) \right ] + \dot f_0t.
\end{equation}
\MDGold{Differentiating} with respect to time Equation \ref{eq:doppler_line_2} gives:
\begin{equation}\label{eq:doppler_line_sd}
    \dot f=-f_0\omega\frac{v}{c} \sin{\beta'_0}\cos{(\lambda_0 - \omega t)} + \dot f_0.
\end{equation}
If we define the parameters $\rho$ and $\dot \rho$ as
\begin{equation}\label{eq:rho}
\rho = f_0\sin\beta'_0\sin(\lambda_0 - \omega t),
\end{equation}
\begin{equation}\label{eq:rhodot}
   \dot\rho=-f_0\omega\sin{\beta'_0}\cos{(\lambda_0 - \omega t)}
\end{equation} 
we can rewrite
\begin{equation}\label{eq:doppler_line_122}
    f= \frac{v}{c}\rho + f_0 + \dot f_0t,
\end{equation}
\begin{equation}\label{eq:doppler_line_sd2}
    \dot f=\frac{v}{c} \dot\rho + \dot f_0.
\end{equation}

\MDGold{At this stage, we observe that, at any observing time time $t = T_{\textrm{obs}}$, Equation \ref{eq:doppler_line_122} depends solely on the sky position through the parameter $\rho$.} \MDG{Moreover, Equation \ref{eq:doppler_line_122} can be simplified further. To this purpose, we point out again that the upper limit of the $\dot f_0$ parameter space for the FH in all-sky searches is $10^{-9}\,{\rm \frac{Hz}{s}}$ and $T_{\textrm{obs}}=\mathcal{O}( 1 \rm\,{yr}) \simeq 3\times 10^{7} \rm\, s$. It follows that the term $\dot f_0 T_{\textrm{obs}}$ contributes at most $\mathcal{O}(10^{-2})$. Given that for current detectors $f_0\ge 20\rm\, Hz$, in the worst possible case $\frac{\dot f_0 T_{\textrm{obs}}}{f_0} = 10^{-3}$ \footnote{It is also straightforward to verify that this approximation holds even for short timescales comparable to the coherence time of the analysis. As an example, if $ T_{\textrm{obs}}=\mathcal{O}(T_{\textrm{coh}}) \simeq \times10^3\rm\,s$ at most (see Table \ref{tab:FH}), the ratio $\frac{\dot f_0 T_{\textrm{obs}}}{f_0}$ is more than 4 orders of magnitude smaller than the case for $T_{\textrm{obs}}=\mathcal{O}( 1 \rm\,{yr})$.}.

Even if we consider an observing time of $T_{\textrm{obs}}=\mathcal{O}( 4 \rm\,{yr}) \simeq 1.2\times 10^{8} \rm\, s$, $\frac{\dot f_0 T_{\textrm{obs}}}{f_0} = 10^{-2}$ at most. The $\dot f_0 T_{\textrm{obs}}$ modulation starts being significant only if $f_0<100\,\rm Hz$, $\dot f  = \mathcal{O}(10^{-9}\,{\rm \frac{Hz}{s}})$, and $T_{\textrm{obs}}=\mathcal{O}( 40 \rm\,{yr}) \simeq 1.2\times 10^{9} \rm\, s$, which is unlikely for current detectors. As a consequence, we deem the modulation induced by the term $\dot f_0 t$ safely negligible for any realistic observing time and we can rewrite Equation \ref{eq:doppler_line_122} as}
\begin{equation}\label{eq:doppler_line_22}
    f= \frac{v}{c}\rho + f_0.
\end{equation}
\MDGold{In conclusion, as demonstrated by Equations \ref{eq:doppler_line_22} and \ref{eq:doppler_line_sd2}, we observe a correlation in the $(f,\rho)$ and $(\dot f,\dot\rho)$ planes at any given $t=T_{\textrm{obs}}$. The question of whether to use any time point within the observation period or a specific value will be addressed in Section \ref{sec:calib_lambda}.}

\MDGold{Consequently, we can conclude that, if candidates are genuinely associated with a source, their frequencies $f = f_{(1)}, f_{(2)}, \ldots$ and sky positions in the $(f,\rho)$ and $(\dot f,\dot\rho)$ planes should follow the patterns defined by Equations \ref{eq:doppler_line_22} and \ref{eq:doppler_line_sd2}. In contrast, spurious candidates, which exhibit random frequency variations not correlated with the sky position, will be scattered randomly across the $(f,\rho)$ and $(\dot f,\dot\rho)$ planes.
Thus, these equations can be leveraged to veto candidates of non-astrophysical origin, selecting only those that align with the expected patterns \cite{Intini}.}

\MDGold{To clarify this point further, we recall that the FH pipeline performs an all-sky search on a discrete sky grid \cite{Astone2014}. Consider a source with a given frequency parameter $f_0$ located at $[\lambda_{0}, \beta_{0}]$. During the all sky search, several positions around $[\lambda_{0}, \beta_{0}]$ are examined. As a result, candidates are found not only at the true position of the source but also at nearby grid points. These candidates are often referred to as "children" of the source and arise due to the discrete nature of the parameter space, where a point on the sky grid may not exactly match the precise location of the source. Consequently, the Doppler correction applied by the pipeline corresponds to that of any sky position on the grid, rather than the exact position of the source. This leads to a residual Doppler shift in the frequency of these candidates relative to $f_0$, because they originate from a sky position that does not exactly match $[\lambda_{0}, \beta_{0}]$. Therefore, these candidates will be distributed around the lines described by Equations \ref{eq:doppler_line_22} and \ref{eq:doppler_line_sd2}. In contrast, candidates not associated with an actual source, whose frequency is not affected by the reconstructed position, will scatter randomly in the $(f,\rho)$ and $(\dot f,\dot\rho)$ planes (see Appendix for further details).}

\subsection{Calibration Parameter $\Lambda$}\label{sec:calib_lambda}
\MDGold{Previous studies \cite{Intini} explored the dependence of Equations \ref{eq:doppler_line_22} and \ref{eq:doppler_line_sd2} on the total observing time to determine whether any arbitrary time could be selected, or if there is an optimal value for $t$. To simplify the notation, we adopt the nomenclature from \cite{Intini} and define the angle $\Lambda = \omega t$ in Equations \ref{eq:rho} and \ref{eq:rhodot}. The study in \cite{Intini} identified a specific value of $\Lambda$, which depends solely on the duration of the observing run and not on the source parameters. This value optimally distributed the candidate parameters along the line with slope $\frac{v}{c}$ (as described in Equations \ref{eq:doppler_line_22} and \ref{eq:doppler_line_sd2}) in the $(f,\rho)$ and $(\dot f,\dot\rho)$ planes, producing a clear linear pattern consistent with these equations. Although \cite{Intini} provided valuable insights into $\Lambda$, we aim to generalize and further clarify the relationship between $\Lambda$ and the duration of the observing run. Specifically, we propose that, by definition, $\Lambda$, which is linked to the observing period, should represent the angle swept by the Earth orbit during the observation time. Therefore, $\Lambda \in [\omega T_{\textrm{start}}, \omega T_{\textrm{stop}}]$, with $T_{\textrm{start}}$ and $T_{\textrm{stop}}$ being the start and end times of the observing run, respectively. Additionally, since the Earth traverses approximately $\simeq 0.98^{\rm o}$ of its orbit per day, and $\Lambda$ is associated with time, it can be effectively related to the total number of days in the observing run.}

As a result, to determine the optimal value of $\Lambda$ \MDGold{and investigate its relationship with the total duration of the observation period}, we consider the full duration of the third observing run (O3) of the LIGO detectors and divide it in shorter segments starting from a duration of 10 days until, with a 10 day increment, we reach the total observing time of 362 days. For each time segment, we software-inject CW signal reported in Table \ref{tab:calibration}
and run the FH pipeline on a 40 deg$^2$ region around the source position and in 1 Hz band containing the injection. After that, since it is unlikely that the injection falls outside this interval, we select all the candidates within the maximum Doppler shift from the source frequency, i.e. $f \in [f_0(1-1\times 10^{-4}), f_0(1+1\times 10^{-4})]$. 
\begin{table}[h!]
\begin{tabular}{c|c}
${f_0}$ [Hz] & 57.4526 \\ 
&\\
$\dot{f}_0$ [Hz/s] & $4.4205\times10^{-10}$\\
&\\
$h_0$ & $5.0\times10^{-25}$\\
&\\
$\lambda\ [^{\circ}]$ & 180.3\\
&\\
$\beta\ [^{\circ}]$ & -58.4\\
\end{tabular}
\caption{Parameters of the CW software injection used in the calibration process.}
\label{tab:calibration}
\end{table}

\begin{figure}
     \centering
     \begin{subfigure}[b]{\linewidth}
         \centering
         \includegraphics[width=\textwidth]{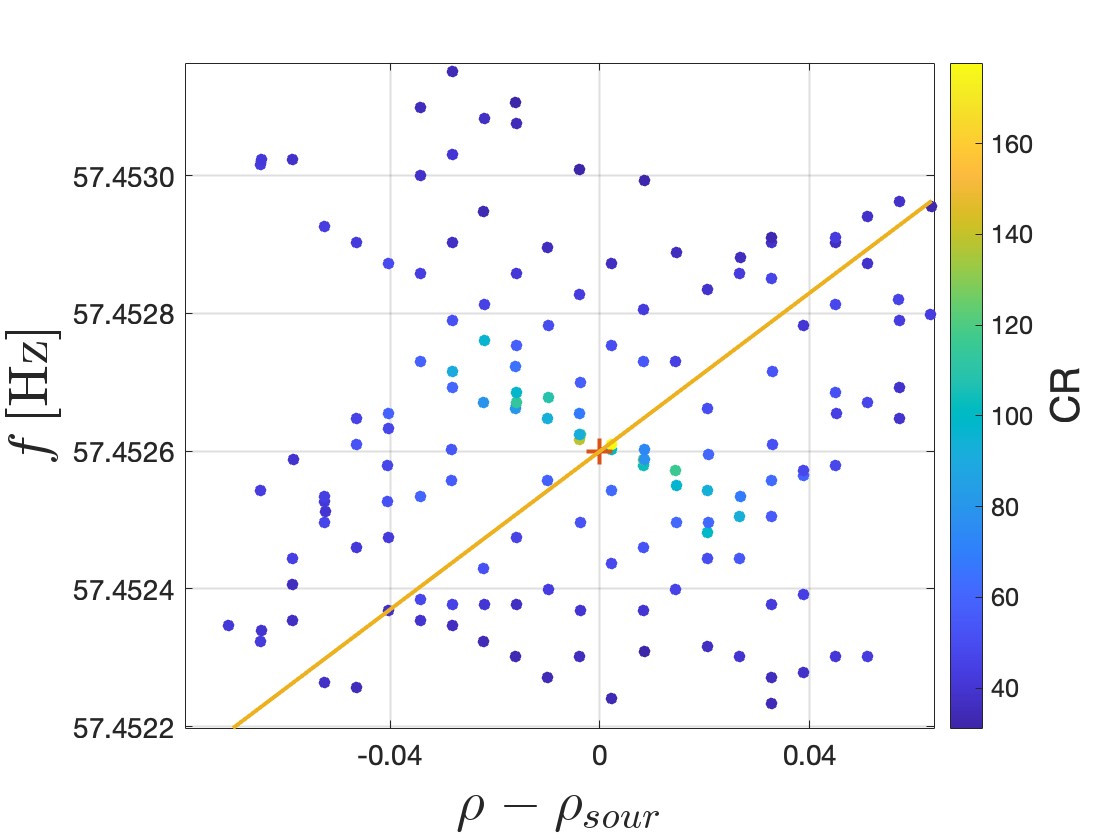}
         \caption{}
         \label{fig:tfevo}
     \end{subfigure}
     \hfill
     \begin{subfigure}[b]{\linewidth}
         \centering
         \includegraphics[width=\textwidth]{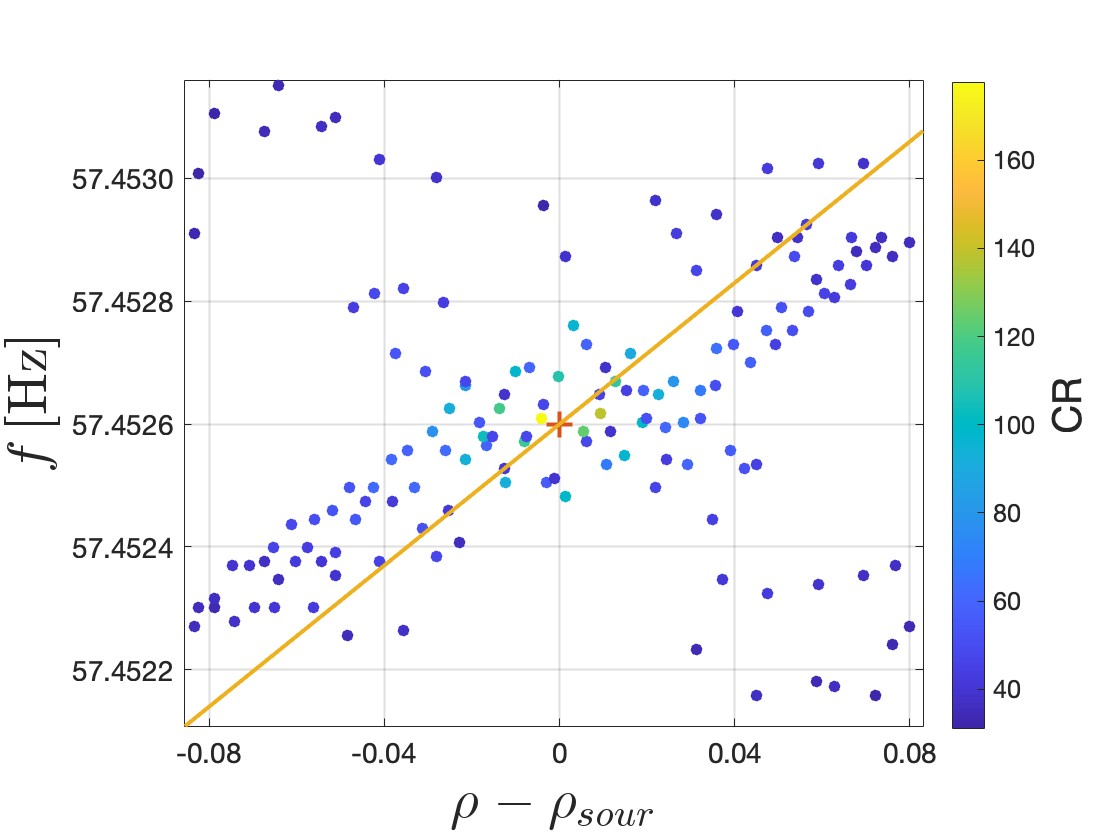}
         \caption{}
         \label{fig:tfevoBNS}
     \end{subfigure}
        \caption{Comparison of the candidates produced by the software injection of Table~\ref{tab:calibration} before (a) and after (b) proper calibration in the $(f,\rho)$ plane. The x-axis has been shifted by the value $\rho_{\textrm{sour}}$ of the injection to be centered at $0$. The red cross marks the injection, whereas the orange line is Equation \ref{eq:doppler_line_22}. The color scale shows the CR of the candidates.}
        \label{fig:before_after}
\end{figure}

At this stage, we retain only the most significant candidates, discarding those with a CR lower than the maximum CR of candidates, which are more likely due to noise, outside the maximum Doppler shift frequency band defined above. \MDGold{Following the prescription of \cite{Intini},} we then calculate the variance, weighted by the CR, of $F(x)=f_0\left[ 1 + \frac{v}{c}\sin\beta'\sin(\lambda - x) \right ]$. \MDGold{At this point, contrary to \cite{Intini} which used a generic time interval,} we variate $x : x \in [\omega T_{\textrm{start}}, \omega T_{\textrm{stop}}]$. The value of $x$ that minimizes $F(x)$ is taken as the calibration value $\Lambda$. Figure \ref{fig:before_after} illustrates the effect of this calibration on a given set of candidates (see also Figure \ref{fig:R3} in the Appendix).

\begin{figure}[h]
    \centering
    \includegraphics[width=\linewidth]{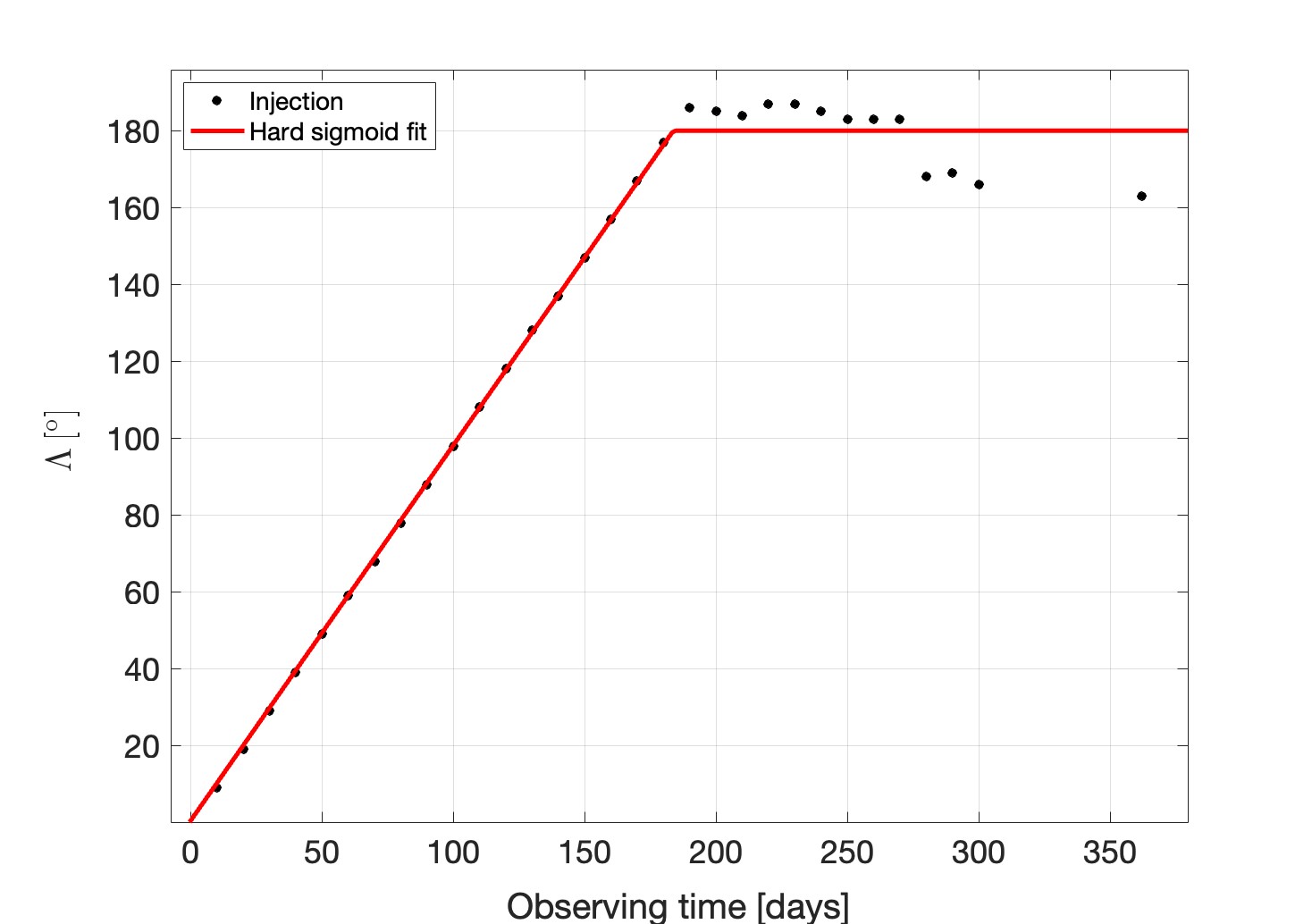}
    \caption{Dependence of $\Lambda$ over the observing time. The hard sigmoid relationship of Equation \ref{eq:sigmoid} is also shown in red.}
    \label{fig:calibration}
\end{figure}

Using this approach, we observe a clear relationship between $\Lambda$ and the duration of the observing run (Figure \ref{fig:calibration}): for observing runs lasting $\le 180$ days, the calibration factor closely aligns with the duration of the observing run in days. For runs with $180 \le T_{\textrm{obs}} \le 365$ days, the calibration factor stabilizes around $\Lambda \simeq 180^{\rm o}$. Additionally, the dependence of $\Lambda$ on time $t$ can be effectively modeled using a hard sigmoid function in the form:

\begin{equation}\label{eq:sigmoid}
    \Lambda(t) = \rm{max} \left [-180; \min\left ( 180; \frac{2t+1}{2}\right)\right].
\end{equation}

This estimate will be confirmed once observing runs with $T_{\textrm{obs}} \ge 365 \rm\, days$ will become available.

\section{Veto procedure}\label{sec:procedure}

Using the method outlined in Section \ref{sec:vetoes},we enhance the veto procedure previously proposed in \cite{Intini} for the FH data analysis pipeline. This improvement focuses on discarding candidates that do not exhibit the specified correlations. These vetoes are applied to all candidates returned by the FH pipeline after detector coincidences and before any post-processing or follow-up procedures.

To detect the linear patterns in the $(f,\rho)$ and $(\dot f, \dot\rho)$ planes \MDGold{at the optimal value of the calibration parameter $\Lambda$}, we use the HT \cite{hough59, hough}, which is well-suited for this task due to its robustness against random patterns that might not be fully removed by other methods \cite{Intini}.

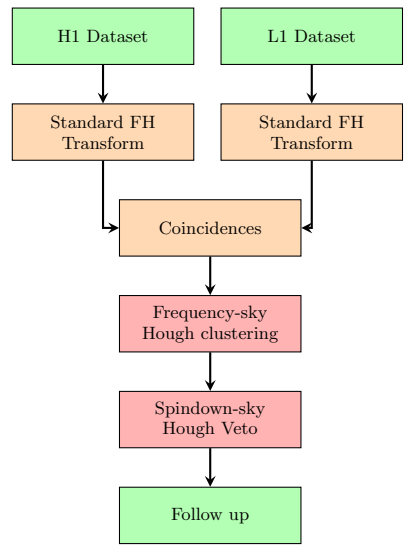
\begin{figure}[t]
\begin{tikzpicture} [node distance=1.7cm]

\node (LLO) [scale=0.75] [detectors] {H1 Dataset};
\node (LLH) [scale=0.75] [detectors, right of=LLO, xshift=2cm] {L1 Dataset};
\node (houghLLO)[scale=0.75]  [process, below of=LLO] {Standard FH Transform};
\node (houghLLH) [scale=0.75] [process, below of=LLH] {Standard FH Transform};
\node (coinci) [scale=0.75] [process, below of=houghLLO, xshift=1.9cm] {Coincidences};
\node (veto1) [scale=0.75] [vetos, below of=coinci] {Frequency-sky Hough clustering};
\node (veto2) [scale=0.75] [vetos, below of=veto1] {Spindown-sky Hough Veto};
\node (followup) [scale=0.75] [detectors, below of=veto2] {Follow up};

\draw [arrow] (LLO) -- (houghLLO);
\draw [arrow] (LLH) -- (houghLLH);
\draw [arrow] (houghLLO) |- (coinci);
\draw [arrow] (houghLLH) |- (coinci);
\draw [arrow] (coinci) -- (veto1);
\draw [arrow] (veto1) -- (veto2);
\draw [arrow] (veto2) -- (followup);

\end{tikzpicture}
\caption{Flowchart for a two detector analysis. The veto chain (red boxes) is added to the standard pipeline after the coincidences between the detectors.}
\label{fig:flowchart}
\end{figure}

The core of the procedure involves incorporating two additional steps into the standard FH pipeline (see Figure \ref{fig:flowchart}), before any follow-up actions are undertaken.

The first step, referred to as the \textit{Frequency-sky Hough clustering} \cite{Intini} from now on, involves searching for the linear pattern in the $(f,\rho)$ parameter space, with $\rho$ calculated using the calibration parameter $\Lambda$ obtained in Section \ref{sec:calib_lambda}. Specifically, the HT maps the candidates from this parameter space to a new $(m,F)$ plane as follows:
\begin{equation}\label{eq:1stveto}
    f=m\rho+F,
\end{equation}
where $m$ must correspond, within a certain error, to $\frac{v}{c}$ (Equation \ref{eq:doppler_line_22}). On the other hand, $F\in[f_{\textrm{min}}, f_{\textrm{max}}]$, where $f_{\textrm{min}}$ and $f_{\textrm{max}}$ are the boundaries of the analyzed frequency band. After the HT, for each value of $F_i$ we select the most significant value of $m_i$. In particular, $m_i$ has to be consistent with $\frac{v}{c}\pm 5\%$. The $5\%$ tolerance on the slope has been found to be \MDGold{the threshold at which the procedure shows a sharp decrease in the number of surviving candidates, dropping suddenly to values near zero. This suggests that the tolerance is more stringent than necessary. For larger values, however, the number of surviving candidates remains nearly constant.} When $m_i$ satisfies the condition defined above, a pattern is considered to be found and a cluster is created. At this point, Equation \ref{eq:1stveto} tells us that $\frac{f-m\rho}{F}=1$, therefore we select all the candidates $j$ in the cluster confined by the two lines 
\begin{equation}
\frac{f_j-m_i\rho_j}{F_i} = 1\pm\delta.    
\end{equation}
As per the selection criteria for the slope,
$\delta$ is a tolerance that we set at $10^{-5}$.

To each cluster obtained in the Frequency-sky Hough clustering, we apply the HT again to map the candidates to a new $(m',\dot F)$ plane following the linear relationship
\begin{equation}
    \dot f=m'\dot{\rho}+\dot F.
\end{equation}
This step is referred to as the \textit{Spindown-sky Hough veto} \cite{Intini}. In this case, $\dot F\in[\dot f_{\textrm{min}}-2f_{\textrm{mean}}\omega\frac{v}{c}, \dot f_{\textrm{max}}+2f_{\textrm{mean}}\omega\frac{v}{c}]$, where $f_{\textrm{mean}}$ is the mean frequency of the cluster. Whereas $m'$ must correspond, within a certain error, to $\frac{v}{c}$ (Equation \ref{eq:doppler_line_sd2}). The selection of $(m',\dot F)$ clusters, accounting for a $5\%$ error on the slope $\frac{v}{c}$, follows a procedure similar to the  Frequency-sky Hough clustering for the same reasons discussed above. This time, for each cluster $i$ we retain only the candidates $j$ such that
\begin{equation}
    |\dot f_j-m_i\dot{\tilde\rho}|\le\dot F + 10\delta\dot f.
\end{equation}

\begin{figure}[b]
    \centering
    \includegraphics[width=\linewidth]{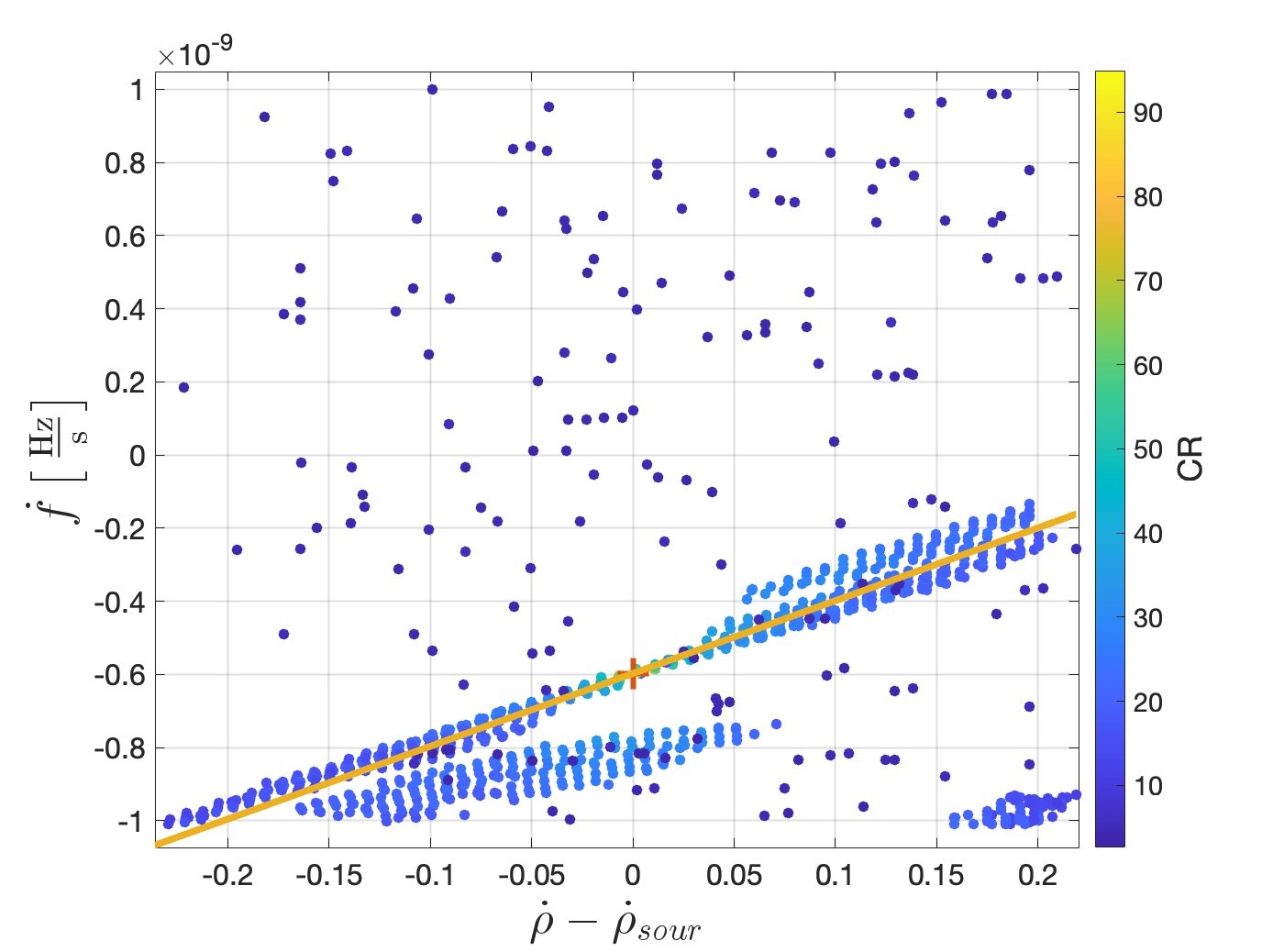}
    \caption{Example of source candidates surviving the Spindown-sky Hough veto. The $(\dot f, \dot \rho)$ pattern is apparent. The orange line shows Equation \ref{eq:doppler_line_sd2}; the red cross marks the injection; the color scale shows the CR of the candidates. The x-axis has been shifted by the source parameter $\dot\rho_{sour}$ to center it at 0.}
    \label{fig:veto_example}
\end{figure}

The outcome of this second step is a set of candidates that have been vetoed. The remaining candidates are then prepared for follow-up analysis.
On average, after considering $\mathcal{O}(10^3)$ software injections uniformly distributed in the parameter space, \MDGold{approximately $84 \%$ of the detector coincident candidates survive after the Frequency-sky Hough clustering.} Of these, $\simeq 53\%$ ($\simeq 44 \%$ of the coincident candidates) are still present after the Spindown-sky Hough veto. These figures are clearly in agreement with \cite{Intini}. An example of surviving candidates after the Spindown-sky Hough veto can be seen in Figure \ref{fig:veto_example}.

In general, we find that the candidates discarded by the veto chain are those which exhibit the lowest statistical significance in the FH pipeline and are mostly caused by noise. On the other hand, the surviving ones are typically those that exhibit the highest CR and are closer to the injection in the parameter space. Nevertheless, we find that some candidates generated by noise can also survive the procedure. This is acceptable, since the aim of the veto chain is to significantly reduce the total number of candidates fed to follow-up procedures, without losing those which are most likely generated by a CW source. The follow-up steps will, in their turn, apply a further selection to properly identify the source.

\section{Procedure efficiency}\label{sec:efficiency}
\begin{figure}
    \centering
    \includegraphics[width=\linewidth]{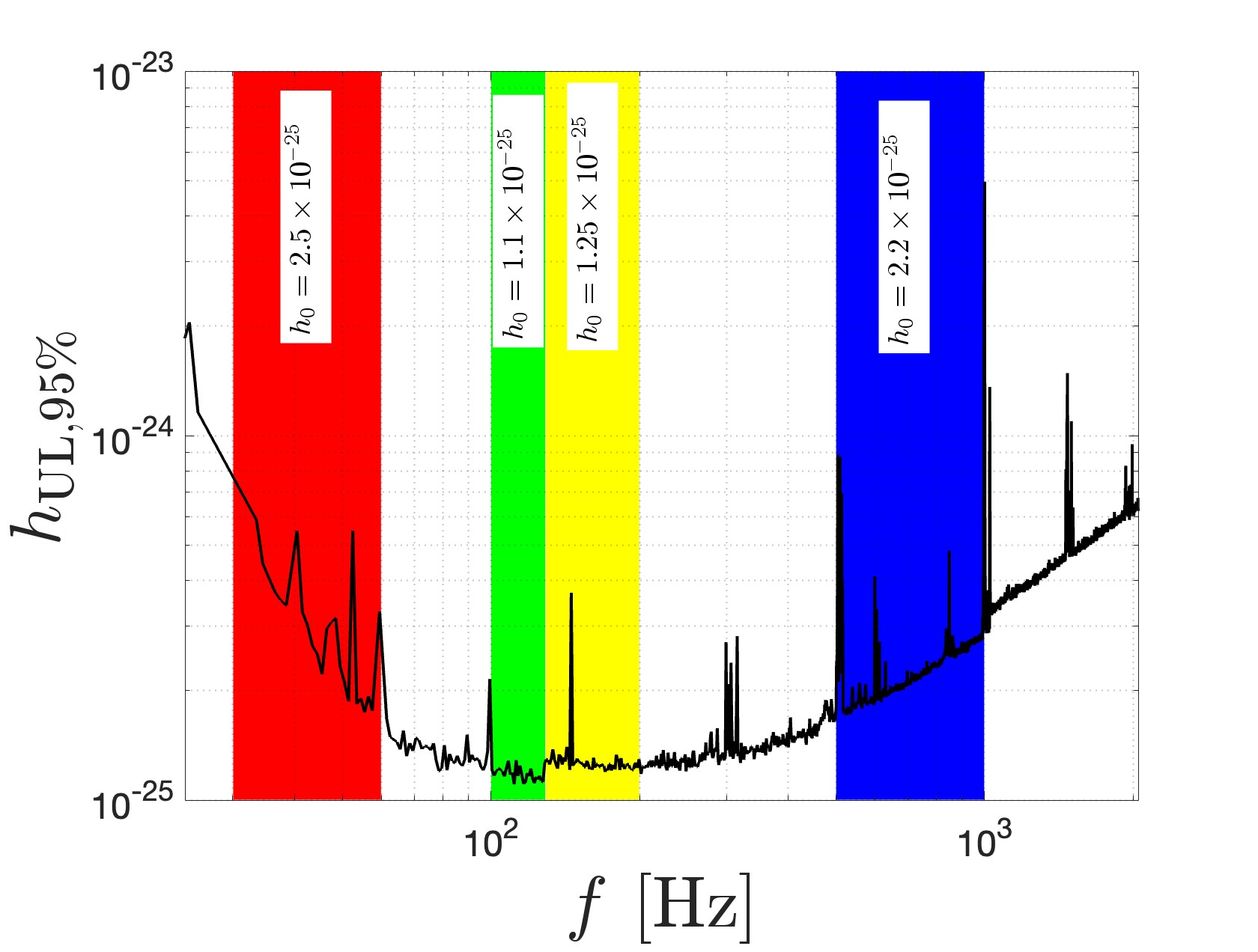}
    \caption{O3 conservative upper limit estimation  for the FH search \cite{CW5}. The colored bands highlight the frequency intervals considered in this work. The $h_0^{up}$ chosen for each band is also shown. The step at $128 \rm\, Hz$ is caused by the different criteria used to select candidates above and below that threshold during the analysis for the upper limits \cite{CW5}.}
    \label{fig:upper_limits}
\end{figure}

To evaluate the efficiency of the procedure, we conduct a MC simulation using software-injected signals in the data from the O3 observing run of the two LIGO detectors. We begin \MDGold{from} the conservative upper limits on $h_0$ as a function of frequency for the FH method, denoted as $h_0^{\textrm{up}}$ (Figure \ref{fig:upper_limits})\MDGold{, which represents the $95\%$ CL on the efficiency of the FH pipeline \cite{CW5}}. From the entire accessible frequency range, we select four sub-bands, as shown in Table \ref{tab:upperlimits} and Figure \ref{fig:upper_limits}. This selection includes the most sensitive frequency ranges of the LIGO detectors  between $100 \rm\, Hz$ and $200\rm\, Hz$, as well as bands where the sensitivity is less optimal.

\begin{table}[t]
\begin{tabular}{c|c}
$\mathbf{f}$ \textbf{[Hz]} & $\mathbf{h_0^{{\rm up}}}$\\
\hline
$[30,60]$ &$2.5\times10^{-25}$\\
&\\
$[100,128]$ &$1.1\times10^{-25}$\\
&\\
$[128,200]$ &$1.25\times10^{-25}$\\
&\\
$[500,1000]$ & $2.2\times10^{-25}$\\
\\
\end{tabular}
\caption{Estimated upper limits for each frequency band (Figure \ref{fig:upper_limits})\cite{o3allsky}. }
\label{tab:upperlimits}
\end{table}

The goal of the MC simulation is to determine, within each frequency band, the value of $h_0$ (denoted as $h_0^{90}$) above which the veto chain successfully recovers at least $90\%$ of the injected signals, compared to those recovered at the amplitude corresponding to $h_0^{\textrm{up}}$. This allows us to infer the amplitude range within which the veto chain ensures a
$90\%$ CL for \MDGold{signal recovery}. \MDGold{In general, at $h_0^{\textrm{up}}$ we expect our procedure to exhibit the same efficiency as the FH pipeline.} For the entire detector frequency range, $h_0^{90}$ is then extrapolated from the results in the selected frequency bands. \MDGold{We would like to emphasize that the goal of the MC simulation is not to make claims about detections, but rather to evaluate the efficiency of the procedure in the presence of a signal. Specifically, the objective is to ensure that the procedure does not completely veto all candidates from a potentially real signal, thereby preserving the efficiency of the FH pipeline. The effects on the false alarm probability of the FH and the robustness of detection can be found in \cite{Intini}.}

The first step of the MC is to inject 100 CW signals, uniformly distributed within each selected frequency band, carefully avoiding known noise lines. These signals have an amplitude corresponding to the $h_0^{\textrm{up}}$ estimated for that band (Table \ref{tab:upperlimits}). The spin-down values of the injections are uniformly sampled within the interval $[-10^{-9};10^{-9}]\rm\, Hz/s$.  The longitude and latitude are uniformly sampled across the intervals $[0;360]\rm\,^o$ and $[-90;90]\rm\,^o$, respectively. The analyzed sky portion is, for $f<500\rm\,Hz$ a $15\rm\, deg^2$ region around the position of the injections. For $f>500\rm\,Hz$, given the higher density of the sky grid at these frequencies \cite{Astone2014}, we reduce the sky region to $7\rm\,deg^2$. Each signal is then analyzed using the FH pipeline, and the candidates are vetoed following the procedure described in Section \ref{sec:procedure}. For each surviving candidate, we calculate the distance from the injected signal in the search parameter space as
\begin{equation}\label{eq:distance2}
\scriptstyle
    d_c=\sqrt{\left ( \frac{f_{c}-f_0}{\delta f} \right )^2 + \left ( \frac{\dot f_{c}-\dot f_0}{\delta \dot f} \right )^2 + \left ( \frac{\lambda_{c}-\lambda_0}{\delta \lambda} \right )^2 + \left ( \frac{\beta_{c}-\beta_0}{\delta \beta} \right )^2},
\end{equation}
where the subscript $c$ identifies the candidate, the subscript $0$ identifies the injection and the parameters space resolutions are defined as in Equation \ref{eq:distance}. For frequencies $f<128\rm\, Hz$, an injection is considered recovered if at least one candidate with $d_c\le 3$ is found after the vetoes. For frequencies $f\ge128\rm\, Hz$, an injection is considered recovered if at least one candidate with $d_c\le 3$ and a CR greater than 5 \cite{CW5} is found after the vetoes. This approach allows us to assess the efficiency at $h_0=h_0^{\textrm{up}}$ in each band, using it as a benchmark.
At this stage, we repeat the entire procedure for progressively decreasing values of $h_0<h_0^{\textrm{up}}$. By doing this, we compare the efficiency at each $h_0$ with the benchmark efficiency inferred at $h_0^{\textrm{up}}$, ultimately identifying the value of $h_0^{90}$. Moreover, we normalize all the obtained efficiencies to 0.95, which is the nominal efficiency associated to the conservative upper limit $h_0^{\textrm{up}}$ that we use as a benchmark.

\subsection{Efficiency and associated error}\label{sec:eff_err}
Given $n$ injections and $k$ recovered events, the efficiency $\epsilon=\frac{k}{n}$ and the associated errors $\sigma_{\epsilon}$ are calculated using the Bayesian approach described in \cite{paterno, ullrich}, i.e. we determine the probability of $\epsilon$ given the measurements of $k$ and $n$:
\begin{equation}\label{eq:bayesian}   P(\epsilon;k,n)=\frac{P(k;\epsilon,n)P(\epsilon;n)}{C},
\end{equation}
where $P(k;\epsilon,n)$ is the probability of $k$ given a certain $\epsilon$ and $n$ is the Binomial
probability; $C = 1/(n+1)$ is a normalization \MDGold{factor ensuring that the total probability is contained in the interval $0\le\epsilon \le 1$} \cite{ullrich}; $P(\epsilon;n)$ is the probability of $\epsilon$ given a value of $n$ that we assume to be uniform between $[0, 1]$. Therefore, Equation \ref{eq:bayesian} can be written as
\begin{equation}\label{eq:probability}
    P(\epsilon;k,n)=\frac{(n+1)!}{k!(n-k)!}\epsilon^k(1-\epsilon)^{n-k}.
\end{equation}
It can be easily verified \cite{paterno,ullrich} that the real efficiency $\epsilon$ is the most probable value (mode) $\hat\epsilon = \epsilon : dP/d\epsilon=0$ of this distribution. In particular,
\begin{equation}\label{eq:mode}
    \frac{dP(\epsilon;k,n)}{d\epsilon} = \frac{\Gamma (n+2)\epsilon^k(1-\epsilon)^{n-k}(n\epsilon - k)}{\Gamma (k+1)\Gamma (n-k+1)\epsilon(\epsilon - 1)}=0 \Leftrightarrow \epsilon = \frac{k}{n},
\end{equation}
where $\Gamma(N)=(N-1)!$ is the gamma function.
On the other hand, the expected value, i.e., the mean, is $\bar\epsilon=\frac{k+1}{n+2}$. \MDGold{For comparison, we recall that the expected value of a Binomial distribution is $\bar k = \epsilon n$.} For large $n$, as in our case, $\hat\epsilon\simeq \bar\epsilon$\MDGold{, meaning that it converges to the efficiency derived from a Binomial distribution} \cite{ullrich}.
The associated uncertainty is \cite{ullrich}
\begin{equation}\label{eq:variance}
    \sigma_{\epsilon} = \sqrt{\mathrm{Var}(\epsilon)} = \sqrt{\bar{\epsilon^2}-\bar\epsilon^2}=\sqrt{ \frac{(k+1)(k+2)}{(n+2)(n+3)}-\frac{(k+1)^2}{(n+2)^2}}.
\end{equation}
It is straightforward to verify that, in the two extreme cases of $k=n$ and $k=0$, for large $n$ the variance is $\lim\limits_{n \to \infty}V(\epsilon) = \frac{1}{n^2}$.

\MDGold{In general, this approach avoids the typical inconsistencies of both the Poissonian and Binomial  distributions. In the former, the error range can extend into unphysical regions where $\epsilon > 1$, , while both distributions assign zero uncertainty if $k=0$ or $k = n$ and/or $n = 1$.}


\subsection{Signal rates with significant candidates vetoed as False Dismissals} 

The injections from the MC simulation (discussed at the beginning of Section \ref{sec:efficiency}) are also used to estimate the signal rates with significant candidates excluded as false rejects. We expect that the probability of discarding a potentially significant source candidate is non zero. For our analysis, and based on common practices for selecting candidates in the FH pipeline \cite{CW5}, we designate a candidate as significant if $d_c\le 3$.

To estimate the rate, we follow this procedure: for each injection, we first count the number of significant candidates immediately after the FH. Next, we assess how many of these candidates remain after the vetoes. If any significant candidate has been discarded, we flag the injection. Finally, for each frequency band and each value of $h_0$, we define our rate as the ratio of flagged injections ($\rm N_{flag}$) to the total number of injections ($\rm N_{tot}$):
\begin{equation}\label{eq:FDR}
\rm{R_{f}}=\frac{N_{flag}}{N_{tot}}.
\end{equation}
This approach allows us to interpret the rate as the probability of having an injection for which at least one significant candidate has been vetoed, even if other candidates remain. The uncertainties over $\rm R_f$ are calculated using the same approach as described in Section \ref{sec:eff_err} by considering, this time, the probability of $\rm R_f$ given the measurements of $\rm N_{flag}$ and $\rm N_{tot}.$

\subsection{Results}
\begin{figure*}
        \centering
        \begin{subfigure}[b]{0.475\textwidth}
            \centering
            \includegraphics[width=\textwidth]{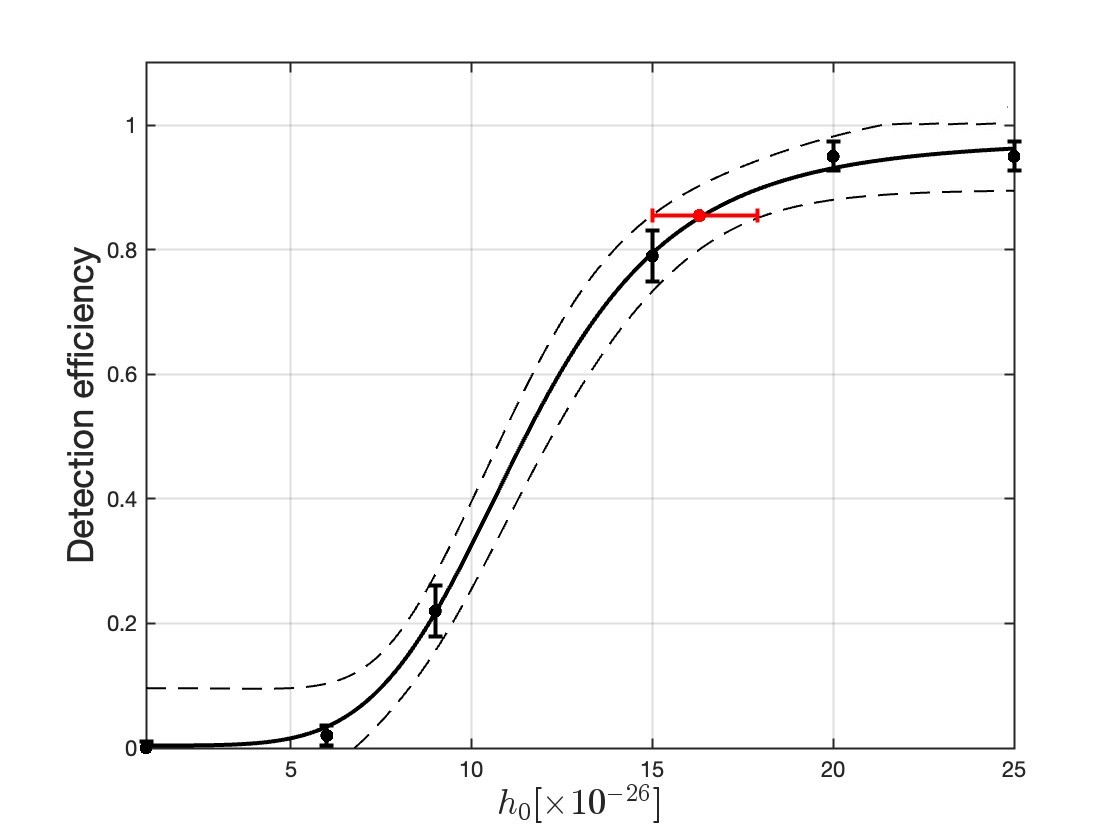}
            \caption[Network2]%
            {{\small $[30,60] \rm\, Hz$}}    
            \label{fig:eff_curves1}
        \end{subfigure}
        \hfill
        \begin{subfigure}[b]{0.475\textwidth}  
            \centering 
            \includegraphics[width=\textwidth]{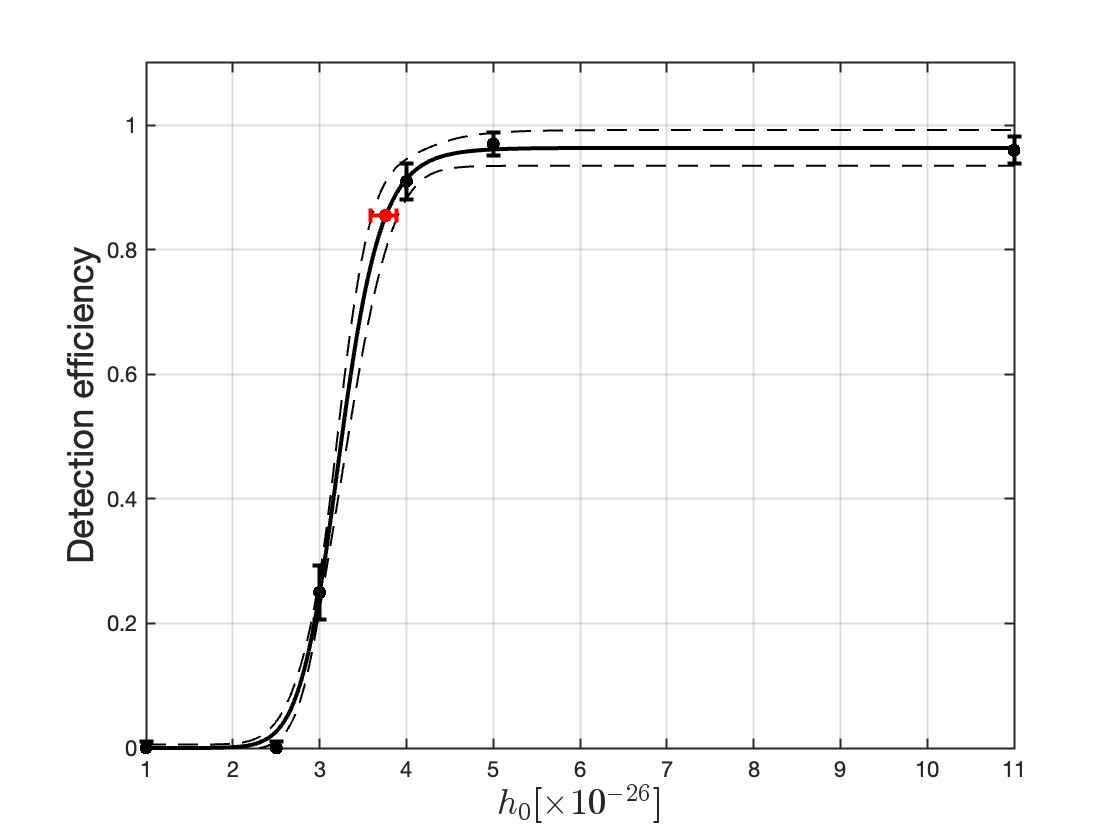}
            \caption[]%
            {{\small $[100,128] \rm\, Hz$}}    
\label{fig:eff_curves2}
        \end{subfigure}
        \vskip\baselineskip
        \begin{subfigure}[b]{0.475\textwidth}   
            \centering 
            \includegraphics[width=\textwidth]{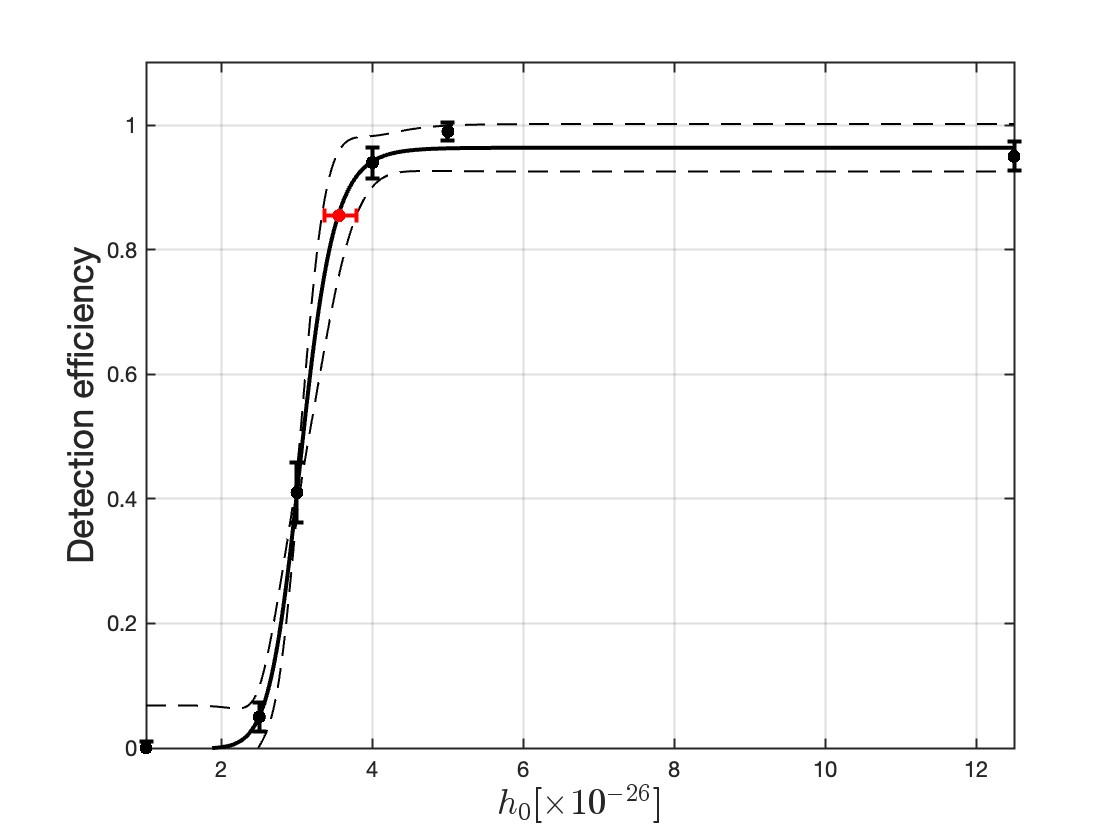}
            \caption[]%
            {{\small $[128,200] \rm\, Hz$}}    
            \label{fig:eff_curves3}
        \end{subfigure}
        \hfill
        \begin{subfigure}[b]{0.475\textwidth}   
            \centering 
            \includegraphics[width=\textwidth]{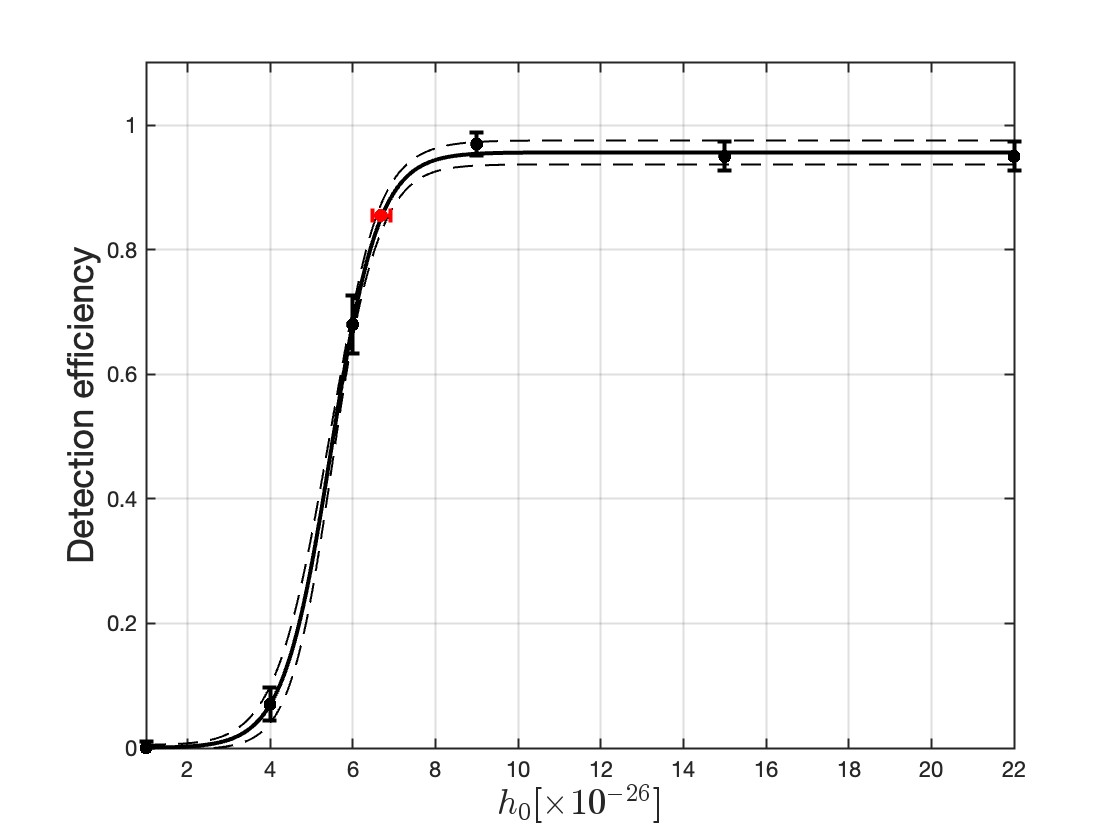}
            \caption[]%
            {{\small $[500,1000] \rm\, Hz$}}    
            \label{fig:eff_curves4}
        \end{subfigure}
        \caption[]
        {Detection efficiencies versus amplitude of the injected signals (black points). The efficiencies are normalized to 0.95 and the error bars are computed according to Equation \ref{eq:variance}. The $90\%$ upper limit on $h_0$ is also shown with the $90\%$ CL (red points). The fit (black lines) is a logistic curve (Equation \ref{eq:logistic}). The $90\%$ CL prediction bounds of the fit are also shown (dashed lines).} 
        \label{fig:eff_curves}
    \end{figure*}

\begin{figure*}
        \centering
        \begin{subfigure}[b]{0.475\textwidth}
            \centering
            \includegraphics[width=\textwidth]{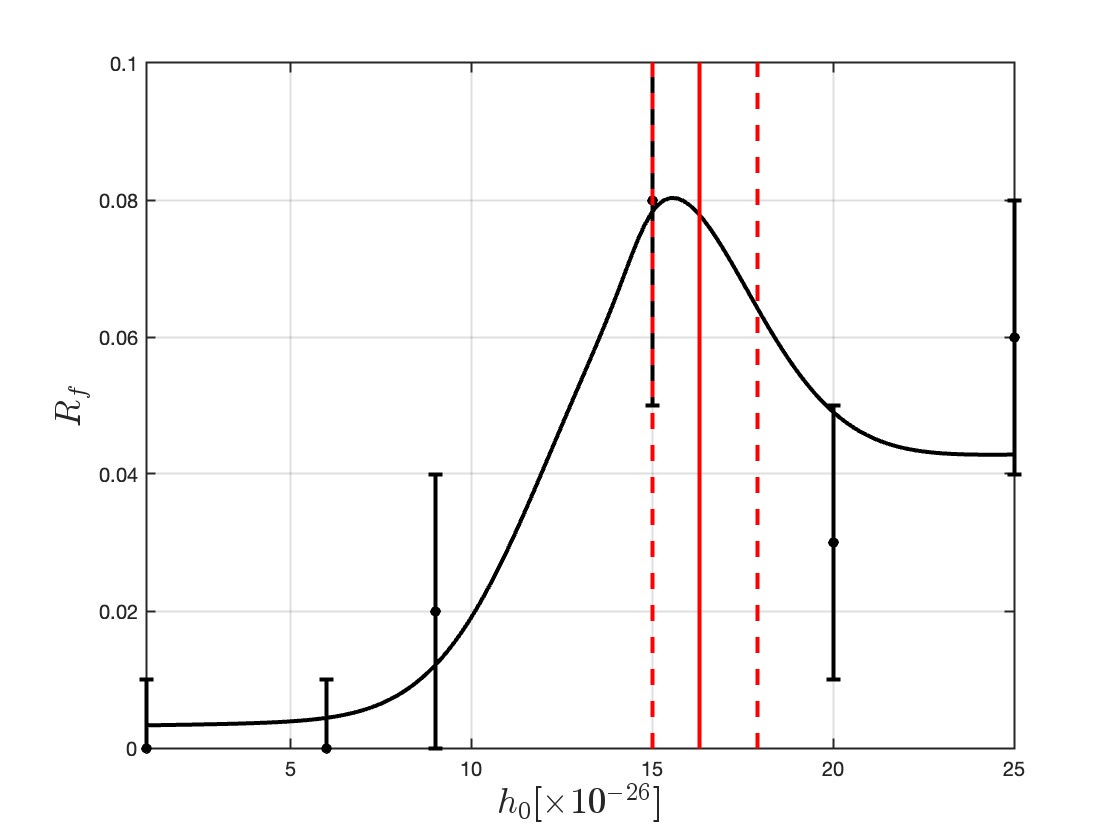}
            \caption[Network2]%
            {{\small $[30,60] \rm\, Hz$}}    
            \label{fig:fdr1}
        \end{subfigure}
        \hfill
        \begin{subfigure}[b]{0.475\textwidth}  
            \centering 
            \includegraphics[width=\textwidth]{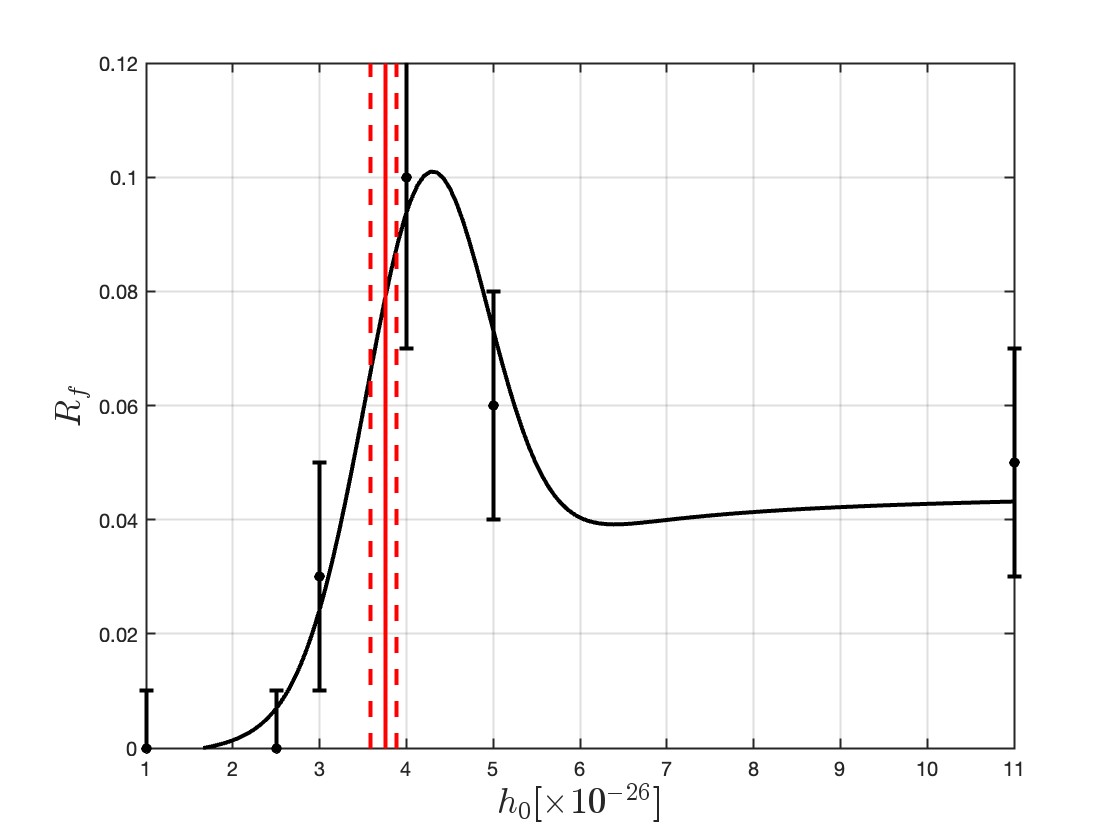}
            \caption[]%
            {{\small $[100,128] \rm\, Hz$}}    
\label{fig:fdr2}
        \end{subfigure}
        \vskip\baselineskip
        \begin{subfigure}[b]{0.475\textwidth}   
            \centering 
            \includegraphics[width=\textwidth]{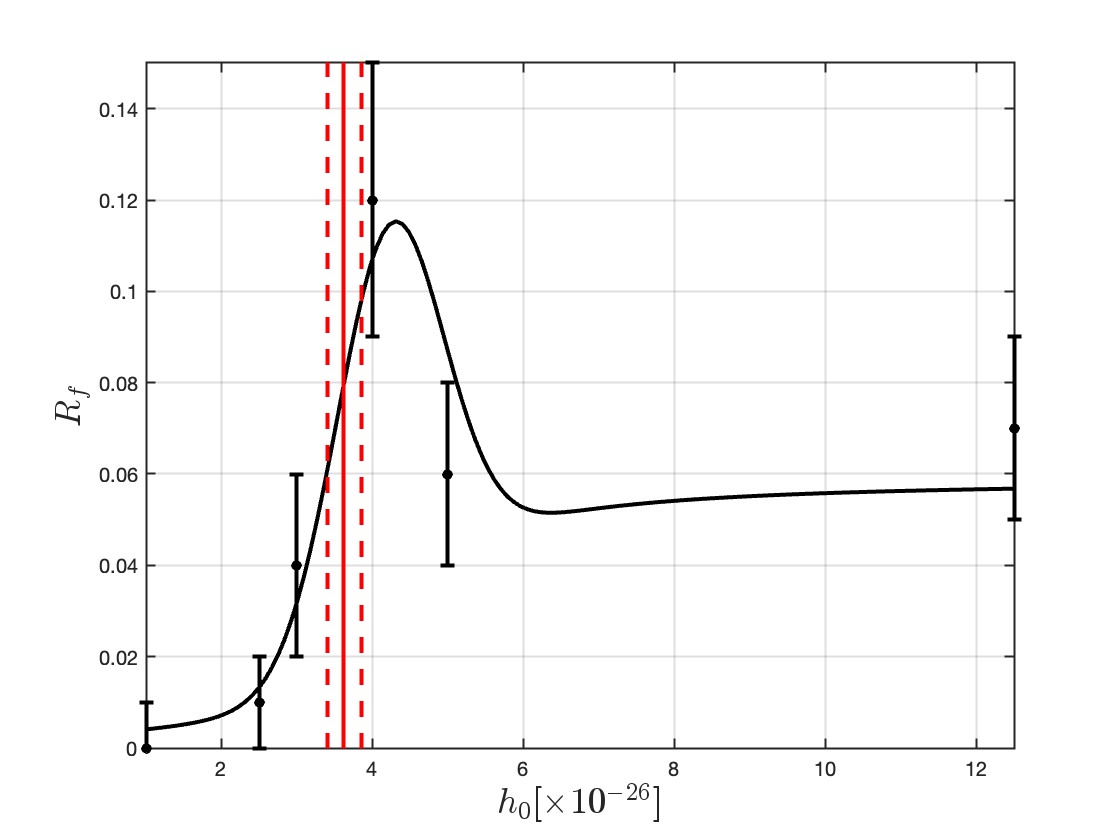}
            \caption[]%
            {{\small $[128,200] \rm\, Hz$}}    
            \label{fig:fdr3}
        \end{subfigure}
        \hfill
        \begin{subfigure}[b]{0.475\textwidth}   
            \centering 
            \includegraphics[width=\textwidth]{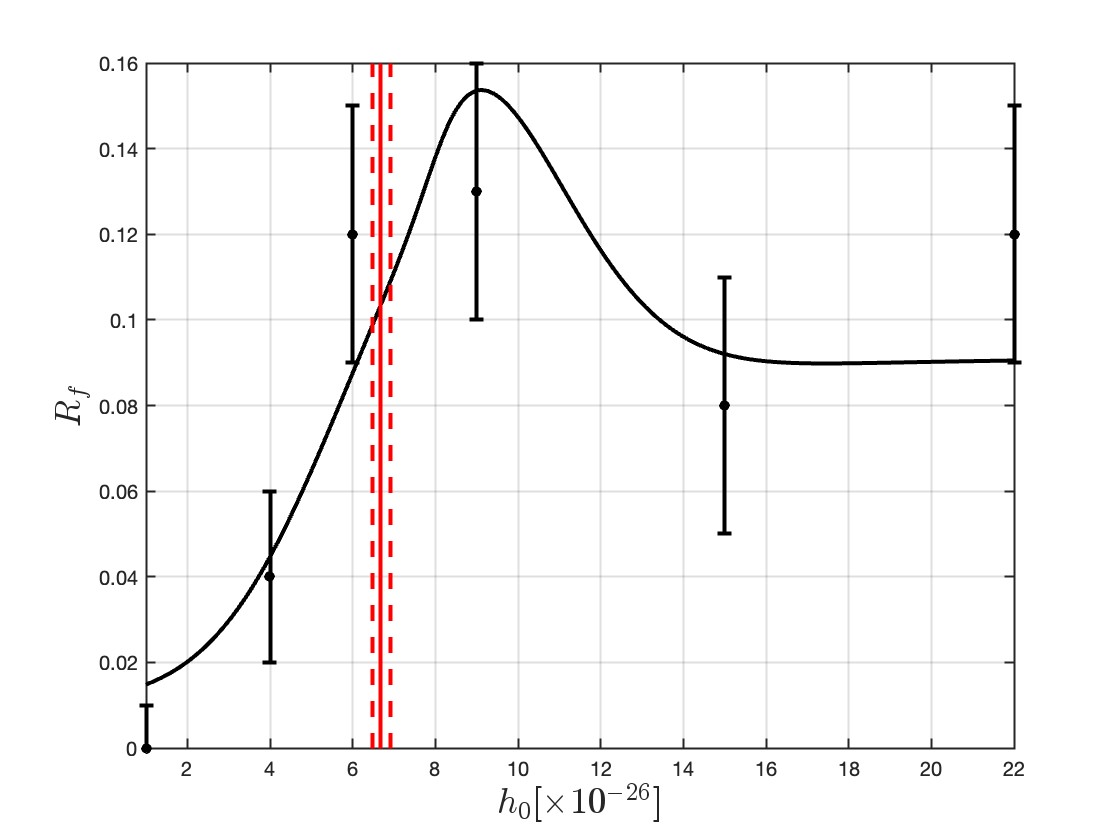}
            \caption[]%
            {{\small $[500,1000] \rm\, Hz$}}    
            \label{fig:fdr4}
        \end{subfigure}
        \caption[]
        {Rate of injections with significant candidates vetoed versus the amplitude of the injected signals (black points). The error bars are calculated using Equation \ref{eq:variance} and, in terms of magnitude, are comparable to those displayed in Figure \ref{fig:eff_curves}. However, they appear different because of the varying scale on the y-axis. The data points are fitted (black curve) with a combination of a Gaussian curve with a sigmoid-like function (Equation \ref{eq:fdr}). Using $h_0^{90\%}$ with its confidence interval (red vertical lines) derived from Figure \ref{fig:eff_curves}, the fit is used to infer $R_f(h_0^{90\%})$ with its confidence interval (Table \ref{tab:upper_lim}).} 
        \label{fig:fdr}
    \end{figure*}

After calculating $\epsilon$ and $\sigma_{\epsilon}$ (Table \ref{tab:efficiencies}), we perform a fit for each selected frequency band. This fit uses a Levenberg-Marquardt algorithm \cite{levenberg} to model the $\epsilon\, \mathrm{versus}\, h_0$ data with a logistic curve
\begin{equation}\label{eq:logistic}
    \epsilon(h_0)=\frac{a}{1+ \exp{\left [ -b(h_0-c)\right ]}},
\end{equation}
with $a,b,c$ being the parameters of the fit. The fitting curve is then used to infer $h_0^{90\%}$ (see Figure \ref{fig:eff_curves} and Table \ref{tab:upper_lim}). To estimate $R_f$ at $h_0^{90\%}$ (Table \ref{tab:upper_lim}), we use a fit to model $R_f$ as a function of $h_0$ with a curve \MDGold{that is the composition of a Gaussian and a sigmoid function.} (Figure \ref{fig:fdr})
\begin{equation*}
    R_f(h_0)=\frac{1}{d}\exp\left [-e(h_0-m)^2\right ]+
\end{equation*}
\begin{equation}\label{eq:fdr}
    +\arctan[f(h_0-m)]+g,
\end{equation}
with $d,e,f,g,m$ being the parameters of the fit. 

In the most sensitive LIGO frequency band considered (i.e., $[100,128]\rm\, Hz$ ), we determine the sensitivity of the procedure at $90\%$ CL to be $h_0^{90\%} = 3.76_{-0.17}^{+0.13}\times 10^{-26}$ with $R_f = 0.079^{+0.009}_{-0.013}$ (Figures \ref{fig:eff_curves2} and \ref{fig:fdr2}). The uncertainty in $h_0^{90\%}$ reflects the $90\%$ CL derived from the fit prediction bounds (as shown in Figure \ref{fig:eff_curves}). The uncertainty on $R_f$ is derived from the confidence interval of $h_0^{90\%}$ (Figure \ref{fig:fdr}). The results for the other frequency bands are reported in Table \ref{tab:upper_lim}.



Since generating an injection set that uniformly covers the entire LIGO-accessible frequency band, from $20 \rm\, Hz$ to $2048 \rm\, Hz$, along with all  the $h_0$ values in the explored amplitude grid would be computationally intensive, we instead extrapolate the upper limits of our procedure to encompass the entire frequency band. This extrapolation employs a method based on the sensitivity estimates presented in \cite{Astone2014}, and utilized in \cite{galactic}.

In particular, the minimum detectable strain amplitude $h_0^{min}$, at a $P$ CL, can be written as \cite{Astone2014}
\begin{equation}\label{eq:extrapolation}
    h_0^{\textrm{min}} = \frac{\mathcal{B}}{N^{1/4}}\sqrt{\frac{S_n(f)}{T_{\textrm{FFT}}}}\sqrt{\rho_{\textrm{CR}}^{\textrm{thr}}-\sqrt{2}\textrm{erfc}^{-1}(2P)},
\end{equation}
where $\rm{erfc}^{-1}$ is the inverse of the complementary error function; $N=\frac{T_{\textrm{obs}}}{T_{\textrm{FFT}}}$ is the effective number of FFTs used in the analysis with $T_{\textrm{FFT}}$ defined in Table \ref{tab:FH}; $P = 0.90$ for our analysis; $S_n(f)$ is the noise strain amplitude of the LIGO Hanford detector, the less sensitive in the network; $\mathcal{B}$ is a parameter that depends on the type of analysis performed and, in our case, is equal to 4.97 (see Appendix B of \cite{Astone2014} and Appendix A of \cite{galactic}); $\rho_{\textrm{CR}}^{\textrm{thr}}$ is the CR threshold used to select candidates. 

\MDGold{However,} as currently stated, Equation \ref{eq:extrapolation} defines the best sensitivity that a given search can achieve when the minimum candidate CR coincides with $\rho_{\textrm{CR}}^{\textrm{thr}}$ \MDGold{, whereas} our goal is to determine the upper limits for the search. To accomplish this, we substitute $\rho_\textrm{CR}^\textrm{thr}$ with the maximum CR ($\rho_\textrm{CR}^{\textrm{MAX}}$) identified within a specific frequency band for injections \MDGold{at} $h_0=h_0^{\textrm{up}}$. Furthermore, since Equation \ref{eq:extrapolation} is affected by the detector sensitivity, we adopt a conservative approach using the detector with the worst sensitivity.

\begin{figure}[htbp]
    \centering
    \includegraphics[width=\linewidth]{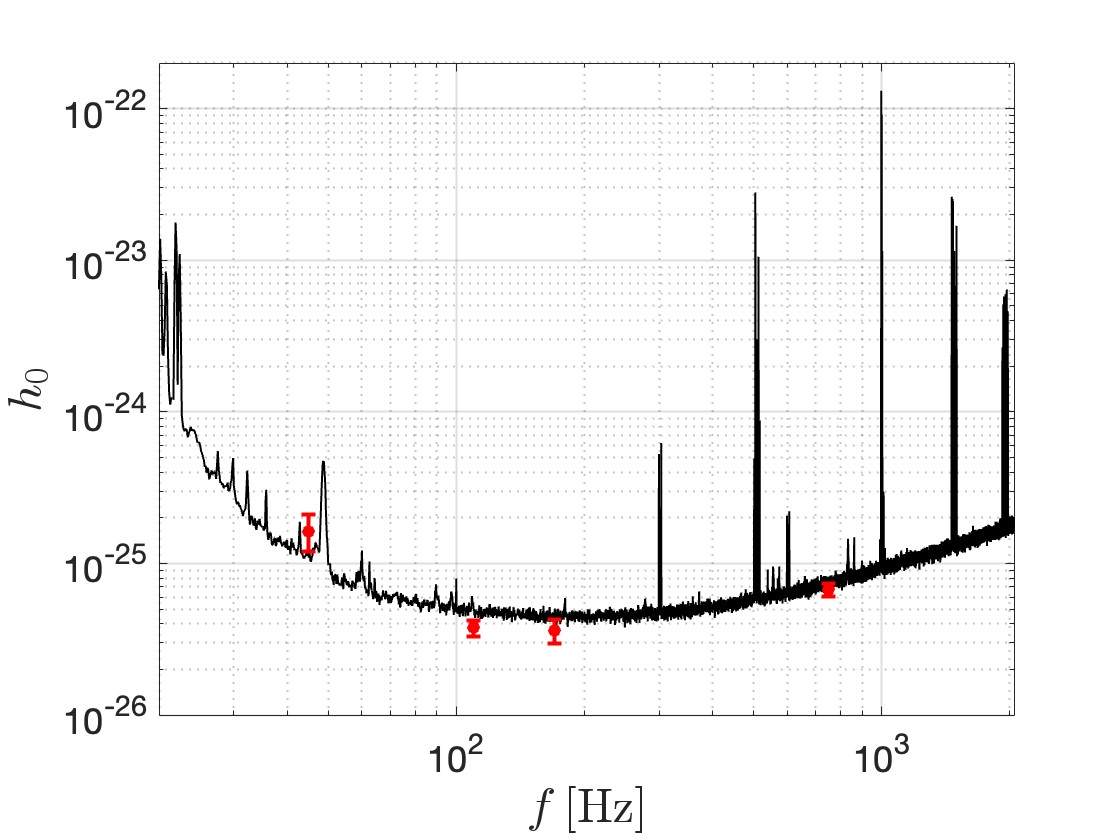}
    \caption{Extrapolation of the sensitivity of the method to the entire LIGO accessible frequency band obtained using Equation \ref{eq:extrapolation}. The red points show the sensitivity values inferred from the MC simulation with the associated error at $3\sigma$ (Table \ref{tab:upper_lim}).}
    \label{fig:interpoaltion}
\end{figure}

Using this approach, we obtain the curve shown in Figure \ref{fig:interpoaltion}. We note that the estimate inferred using Equation \ref{eq:extrapolation} is in agreement with the upper-limit values obtained from the MC simulations.

\section{Conclusions}
\label{sec:conc}
We present a refinement of the methodology initially proposed by \cite{Intini} for a veto chain designed to discard false CW candidates based on correlations between source parameters induced by the Earth Doppler effect. These correlations appear as distinct linear patterns (as described in Equations \ref{eq:doppler_line_22} and \ref{eq:doppler_line_sd2}) within the parameter space. Consequently, we apply the HT to the candidates to identify these patterns. As a result, all candidates that align with these patterns, within a specified tolerance, can be classified as genuine CW candidates.

Using a set of simulated signals uniformly distributed across the parameter space and injected into data from the third observing run of the LIGO detectors, we found that this veto chain effectively discards false CW candidates (see Figure \ref{fig:veto_example}). On average, it retains $\simeq 44\%$ of the \MDGold{candidates identified by the the FH pipeline}, consistently preserving those closest to the injection in parameter space and with the highest CR\MDGold{, thereby confirming the rates reported in \cite{Intini}}.

We also present a more robust explanation of the method calibration (see Figure \ref{fig:before_after}) and demonstrate that the calibration parameter $\Lambda$ is closely related to the duration of the observing run (see Figure \ref{fig:calibration}). Specifically, for observing runs shorter than  180 days, $\Lambda$ corresponds to the number of days in the observing run. Conversely, if $N_{\textrm{days}}>180$, then $\Lambda=180^{o}$ (as shown in Equation \ref{eq:sigmoid}). This estimate will be validated once observing runs longer than one year become available.

Additionally, to determine the upper limits on the amplitude at a $90\%$ CL for the procedure, we conducted a MC simulation consisting of 2400 software injections in O3 data acros four different frequency bands, allowing us to extrapolate the results to the entire accessible bandwdth. 
We found that, in the most sensitive LIGO frequency band considered (i.e., $[100,128]\rm\, Hz$ ), the sensitivity of the procedure at $90\%$ CL is $h_0^{90\%} = 3.76_{-0.17}^{+0.13}\times 10^{-26}$ with $R_f = 0.079^{+0.009}_{-0.013}$ (Figures \ref{fig:eff_curves2} and \ref{fig:fdr2}). The uncertainty in $h_0^{90\%}$ reflects the $90\%$ CL derived from the fit prediction bounds (as shown in Figure \ref{fig:eff_curves}). The uncertainty on $R_f$ is derived from the confidence interval of $h_0^{90\%}$ (Figure \ref{fig:fdr}). The results for the other frequency bands are summarized in Table \ref{tab:upper_lim}. We also found that the probability, over the set of injections, of discarding at least one significant candidate to be consistent, within the uncertainties, for all the frequency bands considered (Tables \ref{tab:efficiencies} and \ref{tab:upper_lim}). 
Finally, combining Equation \ref{eq:extrapolation} \cite{Astone2014, galactic} and the injection set of the MC simulation, we extrapolated the sensitivity estimate of the method to the entire frequency band currently accessible to the LIGO detectors at $90\%$ CL (see Figure \ref{fig:interpoaltion}).

In conclusion, we have demonstrated that this method possesses the necessary foundations to be included in the candidates post processing analyses of the all-sky FH pipeline, thereby enhancing the robustness and statistical significance of follow-up analyses for CW candidates.

Looking ahead, we envision applying this method to other sources of interest for the FH pipeline, such as NSs in binary systems. Searching for a NS in a binary system presents unique challenges due to the distinctive characteristics of the emitted signal compared to an isolated NS. However, if the orbital period of the binary is sufficiently long, the FH pipeline can treat the NS as an isolated source. Consequently, the application of this procedure may also be feasible in this context, thereby increasing the likelihood of retaining only the most significant candidates for robust CW detection.

\acknowledgments
The authors gratefully acknowledge prof. John T. Wheelan and the members of the LVK CW group that, with their precious comments, helped us to improve the presentation of this work. Dr. M. Di Giovanni also acknowledges the contribution of the following Master's students during the preparatory phase of this work: L.
Lunghini, M. Loddo, G. Celani and C. Lauria. Finally, the authors acknowledge the use of the INFN-CNAF computing cluster for the analysis of this work. This material is based upon work supported by NSF's LIGO Laboratory which is a major facility fully funded by the National Science Foundation. This work is also partially supported by ICSC – Centro Nazionale di Ricerca
in High Performance Computing, Big Data and Quantum Computing, funded by
European Union – NextGenerationEU.

\begin{table}[htbp]
\begin{subtable}{\linewidth}
\begin{tabular}{ccc}
\multicolumn{3}{c}{$\bm{[30,60]}$ \textbf{Hz}}\\
\hline
$h_0\times 10^{-26}$ & $\epsilon \pm \sigma_{\epsilon}$& $R_f(h_0)\pm \sigma_{R_f}$ \\
\hline 
$25.0$ & $0.95\pm0.02$ & $0.06\pm 0.02$ \\
$20.0$ & $0.95\pm0.02$ & $0.03\pm 0.02$ \\
$15.0$ & $0.79\pm0.04$ & $0.08\pm 0.03$ \\
$9.0$ & $0.22\pm0.04$ & $0.02\pm 0.02$\\
$6.0$ & $0.02\pm0.02$ & $0.00 + 0.01$\\ 
$1.0$ & $0.00+0.01$&$0.00 + 0.01$\\
\hline
\end{tabular}
\caption{}
\end{subtable}
\newline
\newline
\begin{subtable}{\linewidth}
\begin{tabular}{ccc}
\multicolumn{3}{c}{$\bm{[100,128]}$ \textbf{Hz}}\\
\hline
$h_0\times 10^{-26}$ & $\epsilon \pm \sigma_{\epsilon}$& $R_f(h_0)\pm \sigma_{R_f}$ \\
\hline 
$11.0$ & $0.95\pm0.02$ & $0.05\pm 0.02$\\
$5.0$ & $0.96\pm0.02$ & $0.06\pm 0.02$\\
$4.0$ & $0.90\pm0.03$ & $0.10\pm 0.03$\\
$3.0$ & $0.25\pm0.02$&$0.03\pm 0.02$\\
$2.5$ & $0.01\pm0.01$& $0.00+ 0.01$\\
$1.0$ & $0.00+0.01$ &$0.00 + 0.01$\\
\hline
\end{tabular}
\caption{}
\end{subtable}
\newline
\newline
\begin{subtable}{\linewidth}
\begin{tabular}{ccc}
\multicolumn{3}{c}{\textbf{$\bm{[128,200]}$ Hz}}\\
\hline
$h_0\times 10^{-26}$ & $\epsilon \pm \sigma_{\epsilon}$ & $R_f(h_0)\pm \sigma_{R_f}$ \\
\hline 
$12.5$ & $0.95\pm0.02$ & $0.07\pm 0.02$  \\
$5.0$ & $0.99\pm 0.01$ & $0.06\pm 0.02$  \\
$4.0$ & $0.94\pm 0.03$ & $0.12\pm 0.03$ \\
$3.0$ & $0.41\pm 0.03$ & $0.04\pm 0.02$ \\
$2.5$ & $0.05\pm 0.02$& $0.01\pm 0.01$ \\
$1.0$ & $0.00+0.01$   &$0.00 + 0.01$   \\
\hline
\end{tabular}
\caption{}
\end{subtable}
\newline
\newline
\begin{subtable}{\linewidth}
\begin{tabular}{cccccc}
\multicolumn{3}{c}{\textbf{$\bm{[500,1000]}$ Hz}}\\
\hline
$h_0\times 10^{-26}$ & $\epsilon \pm \sigma_{\epsilon}$ & $R_f(h_0)\pm \sigma_{R_f}$ \\
\hline 
$22.0$ & $0.95\pm0.02$ & $0.12\pm 0.03$  \\
$15.0$ & $0.95\pm0.02$ & $0.08\pm 0.03$  \\
$9.0$ & $0.97\pm 0.02$&$0.13\pm 0.03$   \\
$6.0$ & $0.68\pm 0.05$& $0.12\pm 0.03$  \\
$4.0$ & $0.07\pm 0.03$&$0.04\pm 0.02$  \\
$1.0$ & $0.00+0.01$   &$0.00 + 0.01$   \\
\hline
\end{tabular}
\caption{}
\end{subtable}
\caption{Summary of the detection efficiencies and FDR for each chosen frequency band. All values have been normalized to 0.95.}
\label{tab:efficiencies}
\end{table}


\begin{table}[htbp]
\begin{subtable}{\linewidth}
\begin{tabular}{ccc}
\multicolumn{3}{c}{\textbf{[30,60] Hz}}\\
\hline
$\epsilon^{90}$ & $h_0^{90}\times 10^{-26}$ & $R_f(h_0^{90\%})$\\
\hline 
\\
$0.855$ & $16.3_{-1.3}^{+1.6}$ & $0.078_{-0.017}^{+0.001}$\\
\\
\end{tabular}
\caption{}
\end{subtable}
\newline
\newline
\begin{subtable}{\linewidth}
\begin{tabular}{ccc}
\multicolumn{3}{c}{\textbf{[100,128] Hz}}\\
\hline
$\epsilon^{90}$ & $h_0^{90}\times 10^{-26}$ & $R_f(h_0^{90\%})$\\
\hline 
\\
$0.855$ & $3.76_{-0.17}^{+0.13}$ &$0.079^{+0.009}_{-0.013}$\\
\\
\end{tabular}
\caption{}
\end{subtable}
\newline
\newline
\begin{subtable}{\linewidth}
\begin{tabular}{ccc}
\multicolumn{3}{c}{\textbf{[128,200] Hz}}\\
\hline
$\epsilon^{90}$ & $h_0^{90}\times 10^{-26}$ & $R_f(h_0^{90\%})$\\
\hline 
\\
$0.855$ & $3.62^{+0.23}_{-0.22}$& $0.080^{+0.019}_{-0.039}$\\
\\
\end{tabular}
\caption{}
\end{subtable}
\newline
\newline
\begin{subtable}{\linewidth}
\begin{tabular}{ccc}
\multicolumn{3}{c}{\textbf{[500,1000] Hz}}\\
\hline
$\epsilon^{90}$ & $h_0^{90}\times 10^{-26}$ & $R_f(h_0^{90\%})$\\
\hline 
\\
$0.855$ & $6.68^{+0.23}_{-0.20}$&$0.104_{-0.006}^{+0.004}$\\
\\
\end{tabular}
\caption{}
\end{subtable}
\caption{Summary of the $90\%$  threshold on the efficiency $\epsilon$ and $h_0^{90\%}$ for each frequency band inferred from a logistic curve fit over the data of Table \ref{tab:efficiencies}.}
\label{tab:upper_lim}
\end{table}

\section*{Appendix}
\label{appendix:appendix}
\renewcommand\thefigure{A.\arabic{figure}}    
\setcounter{figure}{0} 

\renewcommand\thetable{A.\arabic{figure}}    
\setcounter{table}{0}  

As discussed in Section \ref{sec:vetoes}, the frequencies of candidates for a potential CW source in all-sky searches exhibit residual Doppler shifts relative to $f_0$, due to the imperfect alignment between the candidate sky positions and the true source position. The FH pipeline corrects for Doppler shifts only at positions on the predefined sky grid, which may slightly deviate from the exact target location \cite{Astone2014}. Consequently, candidates associated with the true source will cluster around the actual source parameters, reflecting the frequency evolution described by Equation \ref{eq:doppler_line_22}. In contrast, candidates unrelated to the true source will display random frequency variations relative to the sky position ($\lambda_0$, $\beta_0$), with no systematic correlation.
\begin{table}[b]
\begin{tabular}{c|c}
${f_0}$ [Hz] & 108.857 \\ 
&\\
$\dot{f}_0$ [Hz/s] & $1.46\times10^{-17}$\\
&\\
$h_0$ & $1.29\times10^{-25}$\\
&\\
$\lambda\ [^{\circ}]$ & 178.3726\\
&\\
$\beta\ [^{\circ}]$ & -33.4366\\
\end{tabular}
\caption{Parameters of the Pulsar 3 hardware injection \cite{GWOSC}. }
\label{tab:calibration2}
\end{table}

\begin{figure}[t]
    \centering
\includegraphics[width=\linewidth]{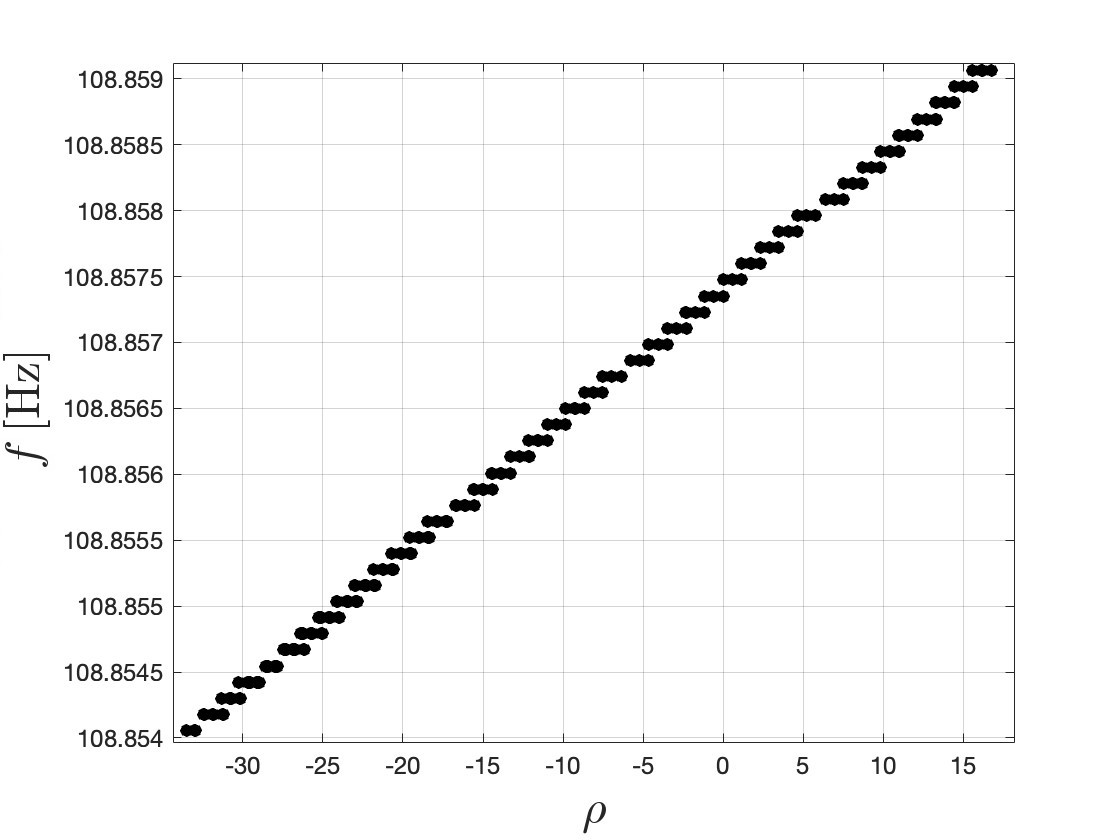}
    \caption{Example of simulated source candidates in the $(f,\rho)$ plane. Being the frequency variation associated with the Doppler modulation, they follow a linear pattern in the parameter space.}
    \label{fig:R1}
\end{figure}

\begin{figure}
    \centering
\includegraphics[width=\linewidth]{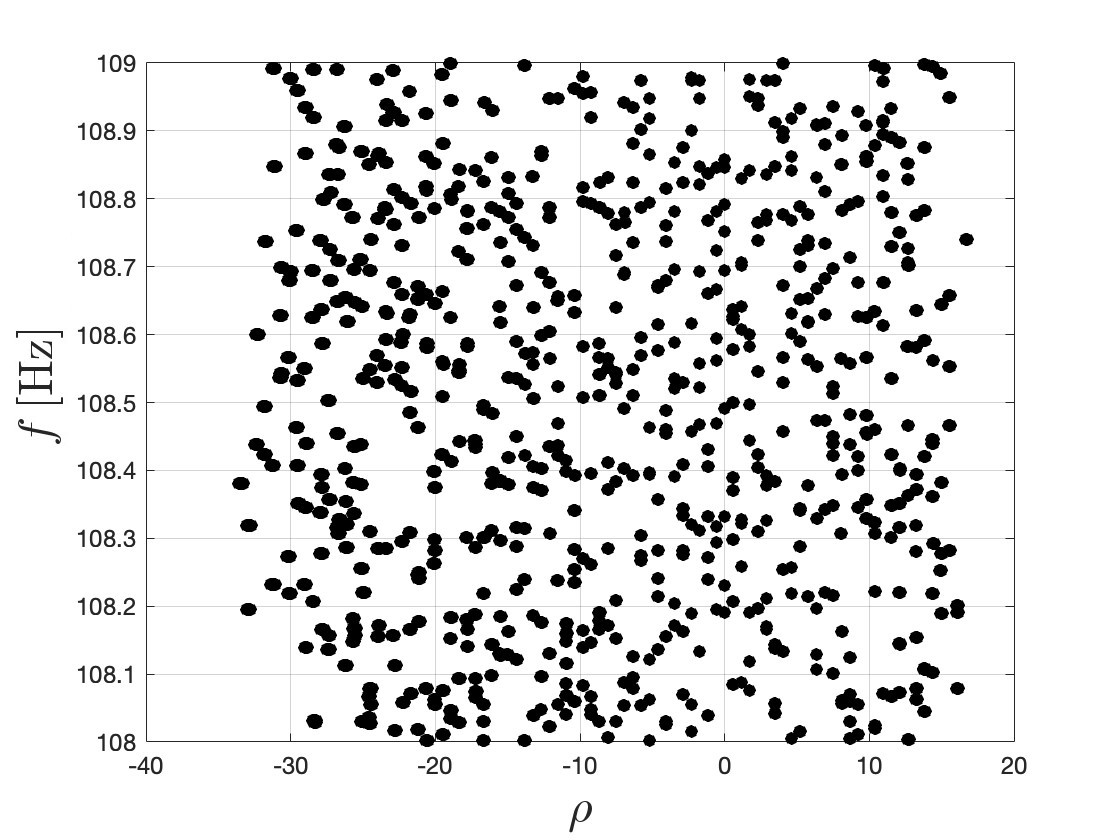}
    \caption{Example of simulated noise candidates in the $(f,\rho)$ plane. Being the frequency variation random, they randomly scatter in the parameter space.}
    \label{fig:R2}
\end{figure}

To clarify this point, we perform a numerical simulation using the parameters from a hardware injection in the LIGO O3 data, specifically \textit{Pulsar 3} (Table \ref{tab:calibration2}). We select a set of points on the sky grid surrounding the true position of the source, calculate the residual Doppler shift at each point, and compute the corresponding parameter $\rho$.
In the $(f,\rho)$ plane, the results show that the frequency of the \textit{children of the source} clusters around the true value $f_0$, with slight variations caused by the Doppler shift (Figure \ref{fig:R1}). The apparent distribution of points along parallel planes, as well as the overlap of some points, arises from the discrete nature of the parameter space. In an ideal scenario with infinitely dense grid sampling, the distribution would rather form a continuous line.
In contrast, candidates unrelated to the source exhibit random variations in both $f$ and $\rho$, as shown in Figure \ref{fig:R2}. Here, we simulate noise-like candidates by randomly sampling an equal number of frequencies, which, by definition, display random frequency variations uncorrelated with the sky position.

\begin{figure*}
        \centering
        \begin{subfigure}[b]{0.475\textwidth}
            \centering           \includegraphics[width=\textwidth]{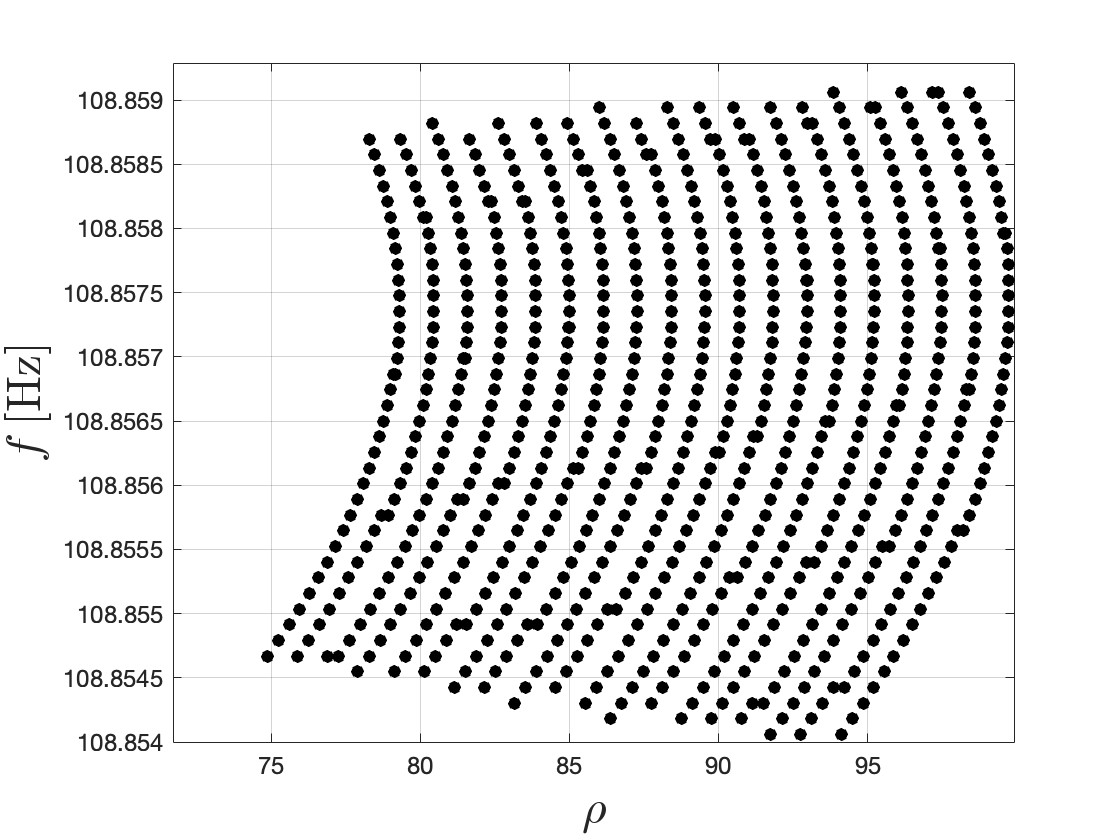}
            \caption{}%
            
            \label{fig:eff_curves1}
        \end{subfigure}
        \hfill
        \begin{subfigure}[b]{0.475\textwidth}  
            \centering 
\includegraphics[width=\textwidth]{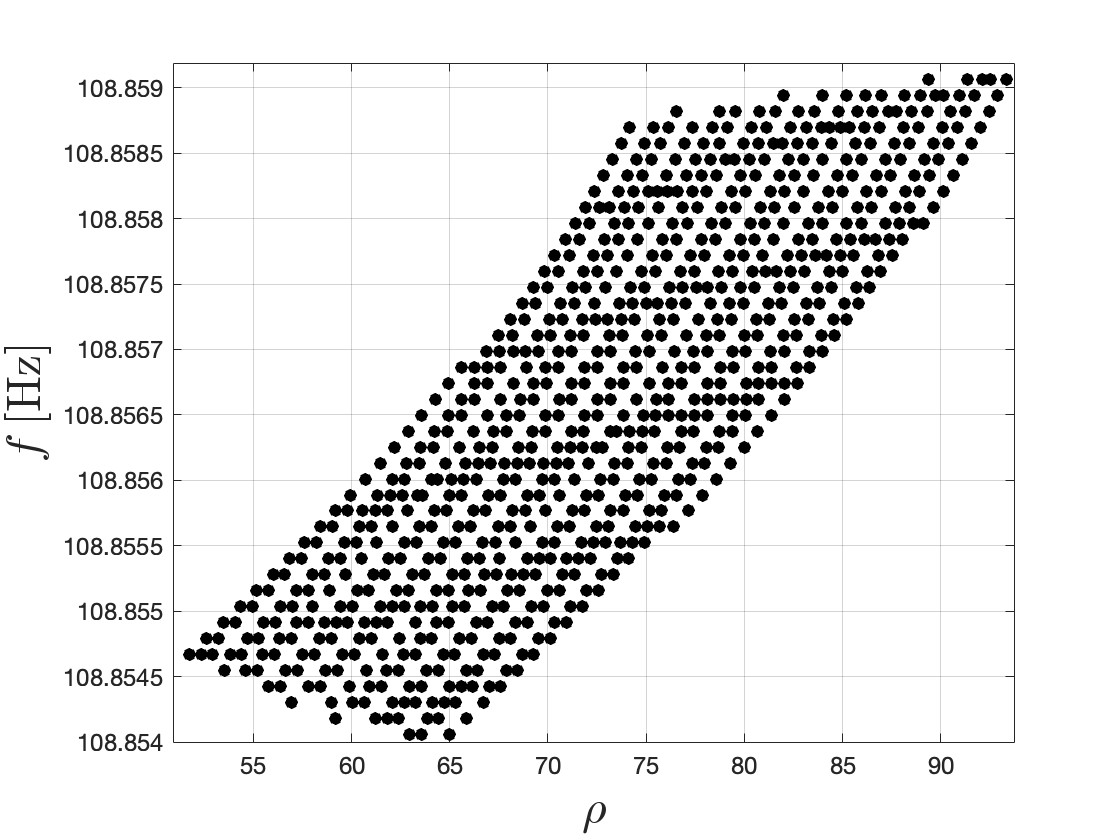}
            \caption[]%
              
\label{fig:eff_curves2}
        \end{subfigure}
        \vskip\baselineskip
        \begin{subfigure}[b]{0.475\textwidth}   
            \centering 
\includegraphics[width=\textwidth]{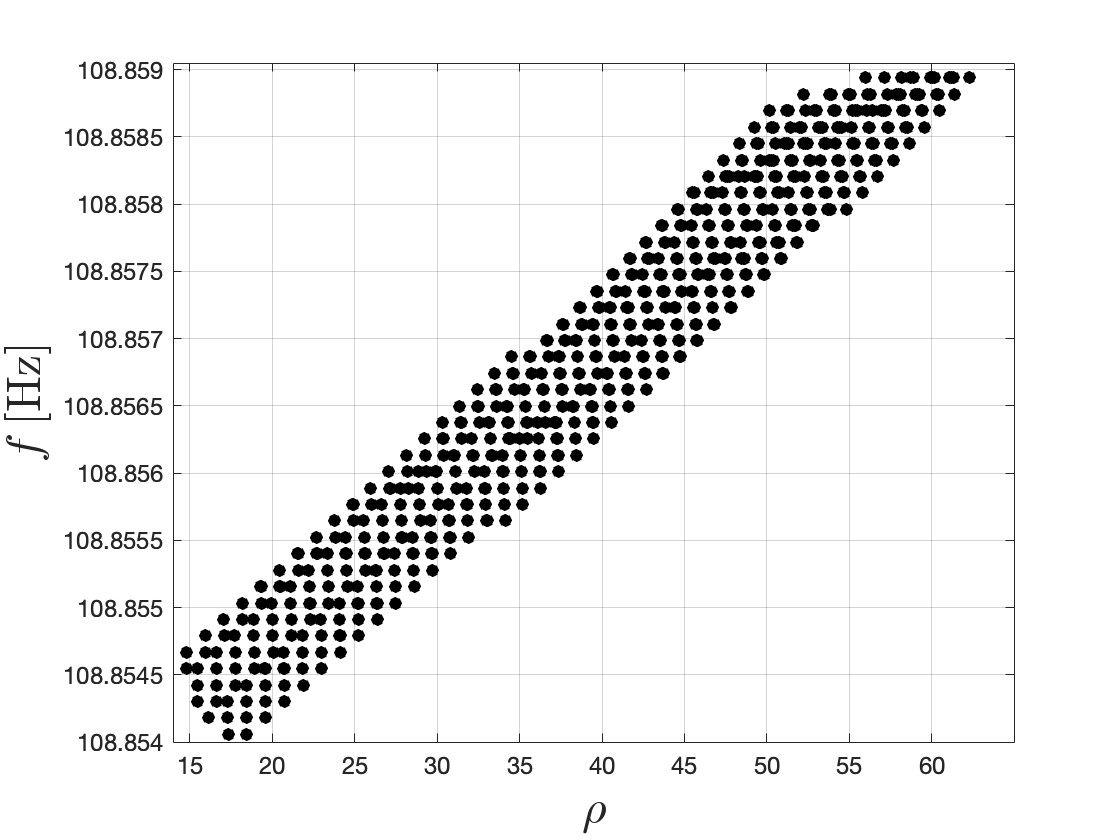}
            \caption[]%
                
            \label{fig:eff_curves3}
        \end{subfigure}
        \hfill
        \begin{subfigure}[b]{0.475\textwidth}   
            \centering 
\includegraphics[width=\textwidth]{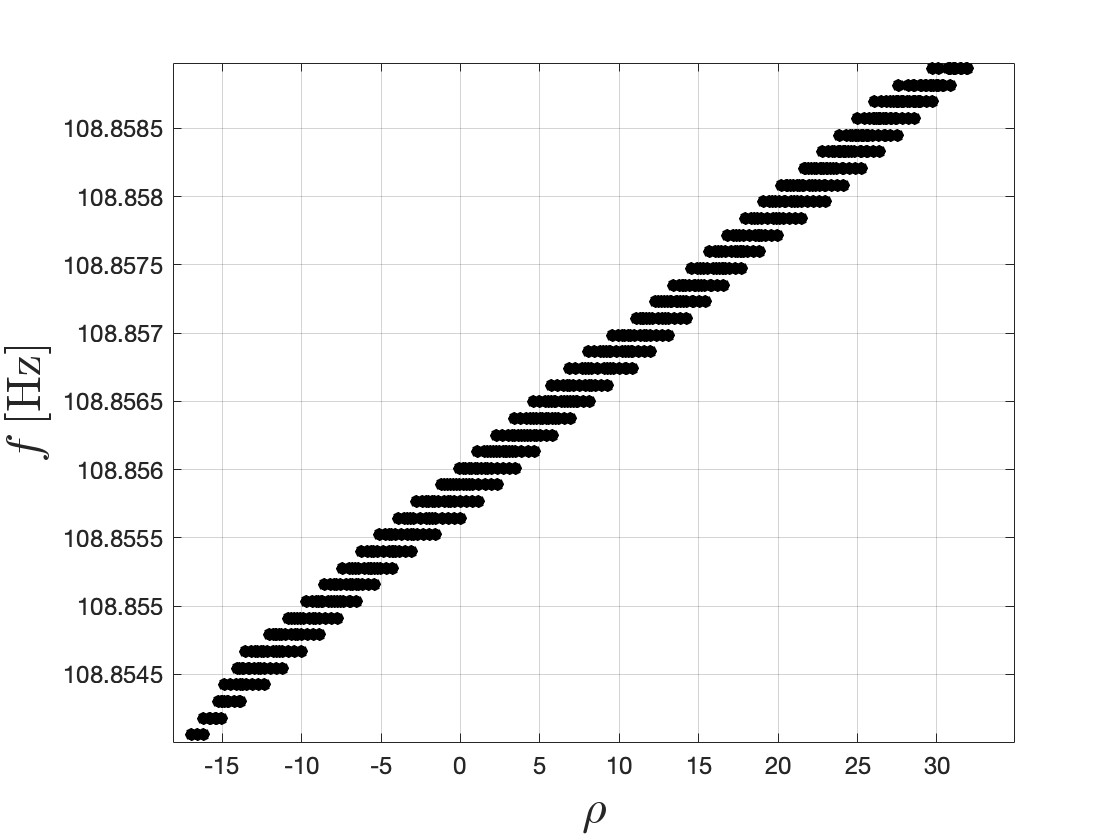}
            \caption[]%
               
            \label{fig:eff_curves4}
        \end{subfigure}
        \caption[]
        {Examples of simulated source candidates with different random values of the calibration parameter. Please note the change of the $\rho$ range, which is caused by different $\Lambda$ values: (a) 0$^o$, (b) 90$^o$, (c) 100$^o$, (d) 160$^o$.} 
        \label{fig:R3}
    \end{figure*}

\MDG{We also point out that Figure \ref{fig:R1} is obtained using the nominal calibration value for the O3 observing run (Figure \ref{fig:calibration}). The effect of the calibration parameter is shown in Figure \ref{fig:R3}, where we can appreciate how, as we approach the nominal value of the calibration parameter for the given observing run, the distribution of points gradually collapses to the well-known linear pattern defined by Equation \ref{eq:doppler_line_22}.}

\bibliography{bibliography}

\end{document}